\documentclass[
               showpacs,preprintnumbers,superscriptaddress,amsmath,amssymb]{revtex4}
\usepackage{graphicx,amssymb,bm}
\usepackage{epsfig}
\usepackage{color}
\usepackage{bbm,verbatim}
\usepackage{subfigure}

\begin{document}

\newcommand{\revlin}[1]{\textcolor{black}{#1}}
\newcommand{\revlina}[1]{\textcolor{black}{#1}}

\newcommand{\dint}{\displaystyle\int}
\newcommand{\doint}{\displaystyle\oint}
\newcommand{\dsum}{\displaystyle\sum}
\newcommand{\bs}[1]{\vec{\boldsymbol{#1}}}
\newcommand{\av}[1]{\langle\mspace{1mu}{#1}\mspace{1mu}\rangle}
\renewcommand{\.}{\boldsymbol{\cdot}}
\newcommand{\x}{\times}
\newcommand{\pdi}[1]{\partial_{#1}}
\newcommand{\td}{\,\mbox{\rm d}}

\newcommand{\be}{\vec{\boldsymbol{e}}}
\newcommand{\bh}{\,\vec{\boldsymbol{h}}}
\newcommand{\bTT}{{\stackrel{\text{\scriptsize$\boldsymbol{\leftrightarrow}$}}
                 {\boldsymbol{\mathbb{T}}}}}
\newcommand{\bTTt}{{\stackrel{\text{\scriptsize$\boldsymbol{\leftrightarrow}$}}
                 {\boldsymbol{\mathbb{T}}}^{\text{\raisebox{-1.8ex}{$t$}}}}}

\newcommand{\bTTo}{{\stackrel{\text{\scriptsize$\boldsymbol{\leftrightarrow}$}}
                 {\boldsymbol{\mathbb{T}}}^{\text{\raisebox{-1.8ex}{$o$}}}}}
\newcommand{\bTToi}{{\stackrel{\text{\scriptsize$\boldsymbol{\leftrightarrow}$}}
                  {\boldsymbol{\mathbb{T}}}^{\text{\raisebox{-1.8ex}{$o$}}}_{\text{inc}}}}
\newcommand{\bTTos}{{\stackrel{\text{\scriptsize$\boldsymbol{\leftrightarrow}$}}
                  {\boldsymbol{\mathbb{T}}}^{\text{\raisebox{-1.8ex}{$o$}}}_{\text{sca}}}}
\newcommand{\bTTox}{{\stackrel{\text{\scriptsize$\boldsymbol{\leftrightarrow}$}}
                  {\boldsymbol{\mathbb{T}}}^{\text{\raisebox{-1.8ex}{$o$}}}_{\text{mix}}}}
\newcommand{\bTToet}{{\stackrel{\text{\scriptsize$\boldsymbol{\leftrightarrow}$}}
                  {\boldsymbol{\mathbb{T}}}^{\text{\raisebox{-1.8ex}{$o$}}}_{\eta}}}

\newcommand{\bTTs}{{\stackrel{\text{\scriptsize$\boldsymbol{\leftrightarrow}$}}
                 {\boldsymbol{\mathbb{T}}}^{\text{\raisebox{-1.8ex}{$s$}}}}}
\newcommand{\bKK}{{\stackrel{\text{\scriptsize$\boldsymbol{\leftrightarrow}$}}
                 {\boldsymbol{\mathbb{K}}}}}
\newcommand{\bKKo}{{\stackrel{\text{\scriptsize$\boldsymbol{\leftrightarrow}$}}
                 {\boldsymbol{\mathbb{K}}}^{\text{\raisebox{-1.8ex}{$o$}}}}}
\newcommand{\bKKox}{{\stackrel{\text{\scriptsize$\boldsymbol{\leftrightarrow}$}}
                 {\boldsymbol{\mathbb{K}}}^{\text{\raisebox{-1.8ex}{$o$}}}_{\text{mix}}}}
\newcommand{\bKKoi}{{\stackrel{\text{\scriptsize$\boldsymbol{\leftrightarrow}$}}
                 {\boldsymbol{\mathbb{K}}}^{\text{\raisebox{-1.8ex}{$o$}}}_{\text{inc}}}}
\newcommand{\bKKos}{{\stackrel{\text{\scriptsize$\boldsymbol{\leftrightarrow}$}}
                 {\boldsymbol{\mathbb{K}}}^{\text{\raisebox{-1.8ex}{$o$}}}_{\text{sca}}}}
\newcommand{\bKKs}{{\stackrel{\text{\scriptsize$\boldsymbol{\leftrightarrow}$}}
                 {\boldsymbol{\mathbb{K}}}^{\text{\raisebox{-1.8ex}{$s$}}}}}
\newcommand{\bKKsx}{{\stackrel{\text{\scriptsize$\boldsymbol{\leftrightarrow}$}}
                 {\boldsymbol{\mathbb{K}}}^{\text{\raisebox{-1.8ex}{$s$}}}_{\text{mix}}}}
\newcommand{\bKKsi}{{\stackrel{\text{\scriptsize$\boldsymbol{\leftrightarrow}$}}
                 {\boldsymbol{\mathbb{K}}}^{\text{\raisebox{-1.8ex}{$s$}}}_{\text{inc}}}}
\newcommand{\bKKss}{{\stackrel{\text{\scriptsize$\boldsymbol{\leftrightarrow}$}}
                 {\boldsymbol{\mathbb{K}}}^{\text{\raisebox{-1.8ex}{$s$}}}_{\text{sca}}}}
\newcommand{\bKKt}{{\stackrel{\text{\scriptsize$\boldsymbol{\leftrightarrow}$}}
                 {\boldsymbol{\mathbb{K}}}^{\text{\raisebox{-1.8ex}{$t$}}}}}
\newcommand{\beps}{{\stackrel{\text{\tiny$\boldsymbol{\leftrightarrow}$}}
                 {\boldsymbol{\mathbb{\epsilon}}}}}
\newcommand{\bI}{\overset{\text{\tiny$\,\boldsymbol{\leftrightarrow}$}}{\boldsymbol{I}}}
\newcommand{\opL}{\boldsymbol{\hat{L}}\,}
\newcommand{\bII}{{\stackrel{\text{\scriptsize$\boldsymbol{\leftrightarrow}$}}
                 {\boldsymbol{\mathbb{I}}}}}
\newcommand{\tc}[1]{{\stackrel{(#1)}{\text{\raisebox{-0.5ex}{$\boldsymbol{\cdot\cdot}$}}}}}
\newcommand{\bOO}[2]{{\stackrel{\text{\scriptsize$\boldsymbol{\leftrightarrow}$}}
                 {\boldsymbol{\mathbb{O}}}_{#1}^{\text{\raisebox{-1.8ex}{$#2$}}}}}
\newcommand{\bnn}[1]{{\stackrel{\text{\tiny$\boldsymbol{\leftrightarrow}$}}
                 {\boldsymbol{n}}^{\text{\raisebox{-0.8ex}{$#1$}}}}}
\newcommand{\bqq}[2]{{\stackrel{\text{\tiny$\boldsymbol{\leftrightarrow}$}}
                 {\boldsymbol{q}}^{\text{\raisebox{-0.8ex}{$#1$}}}_{#2}}}
\newcommand{\bQ}{\overset{\text{\tiny$\boldsymbol{\leftrightarrow}$}}{\boldsymbol{Q}}}

\newcommand{\bM}{{\bs M}}
\newcommand{\bN}{{\bs N}}
\newcommand{\bX}{{\bs X}}
\newcommand{\br}{{\bs r}}
\newcommand{\etheta}{\,{\bs e}_{\theta}}
\newcommand{\ephi}{\,{\bs e}_{\phi}}
\newcommand{\er}{\,{\bs e}_{r}}
\newcommand{\pd}[2]{\frac{\partial{#1}}{\partial{#2}}}
\newcommand{\buu}[1]{{\stackrel{\text{\tiny$\boldsymbol{\leftrightarrow}$}}
                 {\boldsymbol{u}}^{\text{\raisebox{-0.8ex}{$#1$}}}}}

\title{Universal relationships between optical
force/torque and orbital versus spin momentum/angular momentum of light
}

\author{Yikun Jiang}
\affiliation{State Key Laboratory of Surface Physics and Department of Physics, Fudan University, Shanghai 200433, China}
\affiliation{Key Laboratory of Micro and Nano Photonic Structures (Ministry of Education), Fudan University, Shanghai 200433, China}
\author{Huajin Chen}
\affiliation{State Key Laboratory of Surface Physics and Department of Physics, Fudan University, Shanghai 200433, China}
\affiliation{Key Laboratory of Micro and Nano Photonic Structures (Ministry of Education), Fudan University, Shanghai 200433, China}
\author{Jun Chen}
\affiliation{Institute of Theoretical Physics and Collaborative Innovation Center of Extreme Optics, Shanxi University, Shanxi, China }
\author{Jack Ng}
\email[]{jacktfng@hkbu.edu.hk}
\affiliation{Department of Physics and Institute of Computational and Theoretical Studies, Hong Kong Baptist University, Hong Kong}
\author{Zhifang Lin}
\email[]{phlin@fudan.edu.cn}
\affiliation{State Key Laboratory of Surface Physics and Department of Physics, Fudan University, Shanghai 200433, China}
\affiliation{Key Laboratory of Micro and Nano Photonic Structures (Ministry of Education), Fudan University, Shanghai 200433, China}
\affiliation{Collaborative Innovation Center of Advanced Microstructures, Nanjing University, Nanjing 210093, China}

\begin{abstract}
We establish universal relationships between optical force/torque
\revlin{on a general particle}
and different parts of linear and angular momentum (AM) of 
\revlin{generic monochromatic optical field.}
It is rigorously proved that the optical force comes about by the transfer of 
orbital (canonical) optical 
momentum from light to matter, while the other part of optical momentum, known as spin momentum,
does not generate optical force on matter but, instead, stays conserved even when the translational invariance
is broken by putting particles into the optical fields.
On the other hand,
based on a generic multipole theory of optical torque, 
we demonstrate that
the optical torque stems from the transfer 
of the total optical \revlin{AM}, 
including both orbital and spin AM, 
clarifying in generic case the long-standing confusion 
about whether the orbital AM can induce a spinning torque on a general particle in generic optical fields.
\end{abstract}


\maketitle

Light carries linear momentum (LM) and angular momentum (AM)
and thus can exert on any object in the optical field an optical
force and/or an optical torque \cite{maxwell1873,poynting1904,leb1900,nich1903,beth1936}.
The AM of light manifest itself as two constituents, the spin AM and orbital AM \cite{allen1992,allenbook2003,yao2011,humblet1943}.
Analogously, but less obviously, optical LM
falls into two parts as well, viz, the orbital (canonical) part and the spin part, associated, respectively, with
orbital AM and spin AM \cite{belinf1940,
crichton2000,bliokh2013,bliokh2012,bliokh2014,bliokh2014a,beksh2015,berry2009,%
beksh2011rev,bliokh2015,bliokh2016,cameron2012,yao2012,yao2014}.
For a monochromatic optical field,
the period-averaged optical LM density $\av{\bs M}$ read 
\cite{bliokh2013,bliokh2012,bliokh2014,bliokh2014a,beksh2015,berry2009,beksh2011rev,bliokh2015,bliokh2016,cameron2012,yao2012,yao2014}
\begin{subequations}\label{PoPs}
\begin{eqnarray}
\av{\bs M}&\!\!=\!&\!\av{\bs M^s}+\av{\bs M^o}, \label{Mt}\\
\av{\bs M^s}&\!\!=\!&\!\dfrac12\nabla\x\av{\bs S}\label{Ms}\\
\av{\bs M^o}&\!\!=\!&\! 
\dfrac{1}{4\,\omega}\,\text{Im}\,[\varepsilon_0(\nabla\bs E)\.\bs E^* + \mu_0(\nabla \bs H)\.\bs H^*]\label{Mo}
\end{eqnarray}
where $\av{\bs M^s}$ and $\av{\bs M^o}$
are, respectively, known as
 the spin momentum and canonical momentum densities
\cite{bliokh2013,bliokh2014,bliokh2014a,beksh2015,bliokh2015,bliokh2016}. They are
associated with the spin AM density  $\av{\bs S} $ by
Eq.~(\ref{Ms}) and the orbital AM density $\av{\bs L}$ by
\begin{equation}
\av{\bs L}=\bs r\x \av{\bs M^o},
\end{equation}
\end{subequations}
where $\av{\bs S} $ is given by
\cite{berry2009,bliokh2013,bliokh2014,bliokh2014a,beksh2015,bliokh2015,bliokh2016,cohen,barnett2010}
\begin{equation}
\av{\bs S} = \dfrac{\varepsilon_0}{4\,\omega}\,\text{Im}\,[\bs E^*\x\bs E + Z_0^2\,\bs H^*\x\bs H],
\end{equation}
and the total AM density $\av{\bs J}=\av{\bs S}+\av{\bs L}$, with
$Z_0=\sqrt{\mu_0/\varepsilon_0}$ 
being the wave impedance.
For monochromatic optical fields, the time-averaged LM and AM density
can be expressed explicitly in terms of the electric field $\bs E$ and magnetic fields $\bs H$.
The intrinsic nature of the spin AM $\av{\bs S}$  is obvious by its independence of moment arm $\bs r$,
as compared with the extrinsic
orbital counterpart $\av{\bs L}$.

In free-space, the symmetry of translation, rotation and $U(1)$ electromagnetic duality are all
preserved. As a result, the canonical momentum, spin momentum, orbital AM, and spin AM are conserved independently,
as can be seen mathematically from the divergence-free property of the corresponding
time-averaged LM and AM current density tensors (see below). There is no spin-orbit couplings due to the symmetries.

The introduction of any scatterer into the fields, however, breaks all the symmetries, in particular, the $U(1)$ electromagnetic
duality symmetry, due to the absence of magnetic monopoles.
So the LM and AM of light are no longer conserved. The non-conservation 
manifests itself as the optical force and torque acting on the scatterer. 
A question follows naturally whether
one can identify which of the two distinct parts, if not both,
of the optical LM and AM brings about the optical force and torque in generic case, namely,
on arbitrary particle 
in any monochromatic optical fields.

It has been generally believed \cite{oneil2002,garces2003,zhao2007}, at least for some small particles, that
the spin AM causes spinning, a rotation of particle around its center, 
while the orbital AM leads to orbiting, a rotation of particle around some optical field axes.
Theoretically, the spinning can be attributed unambiguously to the optical torque, whereas the orbiting,
involving center-of-mass motion,
can be apparently described by optical force.
In this sense, the optical spinning torque (around particle's center)
comes about due to the transfer of spin AM from light to the particle, while the orbital AM
gives rise to 
optical force. Nevertheless, much less effort has been devoted to examine this belief in generic case.

In this Letter, we establish universal relationships between the different parts of optical LM/AM
and the optical force/torque exerting
on a general particle immersed in generic monochromatic optical fields.
It is rigorously proved that
the period-averaged optical force $\av{\bs F}$ and spinning torque $\av{\bs T}$
can be written as
\begin{subequations}\label{F&T}
\begin{eqnarray}
\av{\bs F}
&=&-c\doint_{S_\infty} \bigl[\langle\bs{M}^o_{\text{tot}}\rangle-\langle\bs M^o_{\text{inc}}\rangle\bigr] \td \sigma, \label{force} \\
&=&-\oint_{S}\av{\bTTo}\.\bs n\td\sigma  \label{force0} \\
\av{\bs T}
&=&-c\doint_{S_\infty} \bigl[\av{\bs J_{\text{tot}}}-\av{\bs J_{\text{inc}}}\bigr]  \td\sigma, \label{torque}\\
&=&-\oint_{S}\bigl[\av{\bKKo}+\av{\bKKs}\bigr]\.\bs n\td\sigma 
 \label{torque0}
\end{eqnarray}
\end{subequations}
where the subscripts ``inc'' and ``tot'' denote, respectively,
the quantities associated with the incident and total fields, or, equivalently,
before and after the particle is put into 
in the field.
The integration in Eqs.~(\ref{force}) and (\ref{torque})
is over a spherical surface $S_\infty$ enclosing the particle with radius $R\to \infty$,
while that in Eqs.~(\ref{force0}) and (\ref{torque0})
is over the outer surface $S$ of the particle.
The dyadics $\av{\bTTo}$, $\av{\bKKs}$, and $\av{\bKKo}$
represent, respectively, the canonical LM current, spin AM current,
and orbital AM current desities. They are given by
\cite{bliokh2013,bliokh2012,bliokh2014,bliokh2014a,beksh2015,berry2009,beksh2011rev,bliokh2015,bliokh2016,cameron2012,yao2012,yao2014}
\begin{subequations}
\begin{eqnarray}
\av{\bTTo}\!\!&=&\!\dfrac{\varepsilon_0c^2}{4\omega}\text{Im}\,
\bigl[(\nabla\bs E)\x\bs B^*+(\nabla\bs B^*)\x\bs E\bigr] \\
\av{\bKKs}\!\!&=&\!\dfrac{\varepsilon_0c^2}{2\omega}\text{Im}\,
\bigl[\bs B \bs E^* +\bs E^* \bs B-(\bs E^*\.\bs B)\bII\bigr]\\
\av{\bKKo}\!\!&=&\!
\dfrac{\varepsilon_0c^2}{4\omega}\text{Im}\,
\bigl[
   \bs B^* \bs E +\bs E \bs B^*
+ (\bs r\x\nabla\bs E)\x\bs B^* \nonumber \\ & & \mspace{25mu} {}
  + (\bs r\x\nabla\bs B^*)\x\bs E \bigr],
\end{eqnarray}
\end{subequations}
with $\bII$ being the unit dyadic.

Equations (\ref{F&T}) are the main results of this Letter. They
contain many salient features of light-induced force and torque.
First, they establish, for the first time and in generic case,
the relationship between the two distinct types of optical LM/AM and their mechanical manifestations, optical force/torque.
It 
\revlin{is suggested}
that the optical force 
can be ascribed totally
to the transfer of the \revlin{canonical momentum} 
from light to matter, while
the optical 
torque results from the \revlin{orbital AM 
as well as the spin AM.} 
Much recent attention has been devoted to explore
the roles of  the two distinct types of LM and AM
on optical force and torque
\cite{yao2012,yao2014,cameron2012,beksh2011rev,bliokh2012,berry2009,bliokh2013,bliokh2014,bliokh2014a,beksh2015,bliokh2015,bliokh2016,saenz2009,
nieto2015a,nieto2015b,pnas2015}.
These studies, however, are confined to small particle (dipolar) limit and no universal
conclusion can be reached.
Eqs.~(\ref{F&T}), which hold generically, 
provide a more transparent and general physical picture for understanding the optical force and torque
as mechanical manifestations of optical LM and AM.
Second, as the optical force is determined by \cite{jackson,zangwill,schwinger}
$\av{\bs F}=-\doint_{S}\av{\bTTt}\.\bs n\td\sigma$,
with $\av{\bTTt}$ denoting \revlin{total LM} 
current density, 
Eq.~(\ref{force0}) implies that $\doint_{S}\av{\bTTs}\.\bs n\td\sigma=0$, with
$\av{\bTTs}=\av{\bTTt}-\av{\bTTo}$ being the spin momentum current. So
the spin momentum of light is still a conserved when a particle is put into the fields,
a signature of its irrelevance to the breaking of translational invariance.
This coincides with our understanding that the intrinsic nature of quantized spin
should not induce force.
The spin momentum 
merely comes into play, 
through its AM counterpart, 
in producing part of the optical torque.
Third, at large distances away from any scatterer,
the surface integral of the normal component of optical canonical LM current $\av{\bTTo}$ and total AM current $\av{\bKKt}=\av{\bKKs}+\av{\bKKo}$ turn out to be that of 
the corresponding LM density $\av{\bs M^o}$ and AM density $\av{\bs J}=\av{\bs L}+\av{\bs S}$, 
except for a multiplicative constant, \textit{i.e.}, the speed  $c$ of light.
Consequently, the optical force (torque) can be described alternatively
by the difference between the surface integrals of orbital LM density (total AM density)
after and before the scatterer is put into the optical fields, provided that
the perturbation to incident optical fields by the introduction of the scatterer is negligible, see, Appendix, for a discussion.
Below we present the 
proof of Eqs.~(\ref{F&T}). 

Let us start with
the optical force, which 
is evaluated
by the surface integration of the Maxwell stress tensor (MST) $\av{\bTT}$ or
LM current density $\av{\bTTt}$ \cite{jackson,zangwill,schwinger},
\begin{equation} \label{emf}
\av{\bs F}=   \oint_S\av{\bTT}\.\bs n  \td \sigma=- \oint_S\av{\bTTt}\.\bs n  \td \sigma,
\end{equation}
where the time-averaged symmetric 
MST $\av{\bTT}$ reads
\begin{equation} \label{Mtensor}
\mspace{-5mu}\begin{array}{ll}
 \av{\bTT} & =
 \dfrac12\text{Re}\bigl[
        \varepsilon_0 \bs E_{\text{tot}} \bs E_{\text{tot}}^* + \mu_0 \bs H_{\text{tot}} \bs H_{\text{tot}}^*  \\
      & \mspace{45mu} {}  - \dfrac12
     (\varepsilon_0 \bs E_{\text{tot}} \. \bs E_{\text{tot}}^*+\mu_0\bs H_{\text{tot}}\.\bs H_{\text{tot}}^*)\bII\,\bigr].
 \end{array}
\end{equation}
The total fields $\bs E_{\text{tot}}$ and $\bs H_{\text{tot}}$ in Eq.~(\ref{Mtensor})
are the sum of the incident fields, $\bs E_{\text{inc}}$ and $\bs H_{\text{inc}}$, and the scattered fields,
$\bs E_{\text{sca}}$ and $\bs H_{\text{sca}}$,
\begin{equation} \label{eht}
\bs E_{\text{tot}} =\bs E_{\text{sca}} +\bs E_{\text{inc}} \quad \text{and}\quad
\bs H_{\text{tot}} =\bs H_{\text{sca}} +\bs H_{\text{inc}}.
\end{equation}
In free-space, the integration 
in Eq.~(\ref{emf}) can be
evaluated over a spherical surface $S_\infty$
with radius $R\to \infty$,
owing to the conservation of optical LM, see, Appendix. 

Decomposition of the total fields Eq.~(\ref{eht}) suggests that
\begin{equation}\label{force3}
\av{\bs F}=\dsum_{\eta}\av {\bs F_\eta} 
                                          ,\qquad \av {\bs F_\eta}=
\displaystyle\oint_{S_\infty}\!\!\av{\bTT_\eta}\.\bs n \td\sigma,
\end{equation}
where the summation runs over $\eta =\text{inc}$, $\text{mix}$, and $\text{sca}$, with
$\av{\bTT_\text{inc}}$ and $\av{\bTT_\text{sca}}$ involving only the incident fields and
scattered fields, respectively, and 
$\av{\bTT_{\text{mix}}}$ collecting all the rest mixed terms \cite{saenz2010,chen2011}, see, also, Appendix.
The incident part $\av{\bTT_\text{inc}}$ gives no net flux 
as understood by the LM conservation law in free-space \cite{born}, see, also, Appendix. 
The mixed and scattering terms yield, respectively, the extinction and recoil forces
\cite{saenz2010,chen2011,chen2015,bohren}, see, also, Appendix.

A generic monochromatic field  incident on a particle
can be written
in the form (see, Appendix) 
\begin{equation} \label{ehinc}
\begin{array}{ll}
\bs E_{\text{inc}}=\doint_{4\pi} \be_{\bs u} \,e^{i k \bs u\.\bs r}\,\td\Omega_u,   \\[3mm] 
\bs H_{\text{inc}}=\dfrac1{Z_0}\doint_{4\pi} \bh_{\bs u} \,e^{i k \bs u\.\bs r}\,\td\Omega_u,
\end{array}
\end{equation}
where
$\bs u$ is the real unit vector denoting the direction of wave vector,
and $\be_{\bs u}$ and $\bh_{\bs u}$ depend on $\bs u$ but not on $\bs r$.
The integration 
is over the unit sphere of directions of wave vectors $k\bs u$. 

The asymptotic behavior of the scattered fields from the particle at large distance 
is characterized by \cite{born} 
\begin{equation} \label{ehs}
\begin{array}{ll}
\bs E_{\text{sca}}= \left(\bs a_{\bs n}+\dfrac{\bs n\,\alpha_{\bs n}}{kr}\right)\dfrac{e^{ikr}}{kr}, \\[2mm] 
\bs H_{\text{sca}}= \dfrac1{Z_0}\left(\bs b_{\bs n}+\dfrac{\bs n\,\beta_{\bs n}}{kr}\right) \dfrac{e^{ikr}}{kr},
\end{array}
\end{equation}
where $\bs n=\bs r/r$ denotes the radial direction.
The vectors $\bs a_{\bs n}$ and $\bs b_{\bs n}$ characterize the transverse
amplitudes of the scattered fields,
while the scalars $\alpha_{\bs n}$ and $\beta_{\bs n}$ denote the longitudinal field components.
They are all
dependent on $\bs n$ but independent of $r=|\bs r|$.

Plugging Eqs.~(\ref{ehinc}) and (\ref{ehs}) into Eq. (\ref{force3}) and using
Jones' lemma \cite{born,saenz2010}, viz,
\begin{equation} \label{jones}
\begin{array}{ll}
\displaystyle\lim_{kR\to\infty}\frac1{kR}\oint_{S_{\infty}}
\bs G(\bs n) e^{-i k (\bs u\.\bs n)R}\,\td\sigma \\[4mm] \mspace{70mu}
\sim
\dfrac{2\pi i}{k^2} \bigl[\, \bs G(\bs u)e^{-ikR} - \bs G(-\bs u)e^{ikR}\,\bigr],
\end{array}
\end{equation}
with $\bs G(\bs n)$ being an arbitrary function of $\bs n$ and $\bs u$ is a real unit vector,
one obtains the extinction and recoil forces (see 
Appendix for details) 
\begin{equation}\label{forcet}
\begin{array}{ll}
\av{\bs F_{\text{mix}}}
&\!\!
=\dfrac{2\pi\varepsilon_0}{k^2}\,\text{Im}\displaystyle \oint_{4\pi}(\be^{\,*}_{ \bs n}\.\bs a_{\bs n})
               \,\bs n \td\Omega_n, \\[3mm]
\av{\bs F_{\text{sca}}}
&\!\!
=-\dfrac {\varepsilon_0}{2k^2}\displaystyle\oint_{4\pi} \bigl|\bs a_{\bs n}\bigr|^2\bs n\td\Omega_n.
\end{array}
\end{equation}

We next turn to the right hand side of Eq.~(\ref{force}).
The decomposition of fields Eq.~(\ref{eht}) implies that
\begin{equation}\label{Moab}
\av{\bs{M}^o_{\text{tot}}}-\av{\bs M^o_{\text{inc}}}=
\av{\bs{M}^o_{\text{mix}}}+\av{\bs M^o_{\text{sca}}},
\end{equation}
with $\av{\bs{M}^o_\eta}$ defined similarly to Eq. (\ref{force3}), see, Appendix. With
Jones�� lemma, Eq. (\ref{jones}), it is not hard to derive, see, Appendix, 
\begin{equation}\begin{array}{ll}
\doint_{S_\infty}\langle\bs{M}^o_{\text{mix}}\rangle\td\sigma
=-\dfrac{2\pi\varepsilon_0}{k^2c}\,\text{Im}\displaystyle \oint_{4\pi}(\be^{\,*}_{ \bs n}\.\bs a_{\bs n})
               \,\bs n \td\Omega_n, \\[3mm]
\doint_{S_\infty}\langle\bs{M}^o_{\text{sca}}\rangle\td\sigma
=\dfrac{\varepsilon_0}{2k^2c}\displaystyle\oint_{4\pi} \bigl|\bs a_{\bs n}\bigr|^2\bs n\td\Omega_n,
\end{array}
\end{equation}
which constitutes the proof of Eq.~(\ref{force}).

Similarly, based on Eq.~(\ref{eht}), one writes 
\begin{equation}\label{Toab}
\av{\bTTo}= \av{\bTToi}+ \av{\bTTox}+\av{\bTTos},
\end{equation}
where $\av{\bTToet}$ is defined similarly to $\av{\bTT_{\eta}}$ in Eq.~(\ref{force3}), see, Appendix. 
it follows directly from Jones' lemma 
that, see, Appendix,
\begin{equation}
\begin{array}{ll}
\doint_{S_\infty}\!\!\!
\av{\bTTox}\.\bs n\td\sigma=c
\doint_{S_\infty}\langle\bs{M}^o_{\text{mix}}\rangle\td\sigma
\\[3mm]
\doint_{S_\infty}\!\!\!
\av{\bTTos}\.\bs n\td\sigma=c
\doint_{S_\infty}\langle\bs{M}^o_{\text{sca}}\rangle\td\sigma.
\end{array}
\end{equation}
Noticing that the integral of $\av{\bTToi}$ vanishes identically,
one concludes the proof of Eq.~(\ref{force0}).

The proof of Eq.~(\ref{torque}) turns out much more complicated due to the presence of the moment arm $\bs r$ in
the orbital AM $\av{\bs L}$. Starting, once again, with the MST, 
one has the spinning optical torque \cite{jackson,zangwill,schwinger}
\begin{equation} \label{emt}
\av{\bs T}=\dsum_{\eta}\av{\bs T_{\eta}} 
 ,\qquad \av {\bs T_\eta}=-
\oint_{S_\infty}\av{\bKK_{\eta}}\. \bs n \td \sigma,
\end{equation}
where $\av{\bKK_{\eta}}=-\bs r\x\av{\bTT_\eta}$, with the subscript $\eta =\text{inc}$, $\text{mix}$, and $\text{sca}$,
similarly to Eq.~(\ref{force3}).
The incident term $\av{\bs T_\text{inc}}$ vanishes identically as understood by the AM conservation law.
The mixed and scattering terms, $\av{\bs T_\text{mix}}$ and $\av{\bs T_\text{sca}}$
 are called the extinction and recoil torques \cite{saenz2010,nieto2015a,nieto2015b,chen2011,chen2015,bohren}, respectively.

With Eqs.~(\ref{eht}), (\ref{ehinc}), and (\ref{ehs}), one has
the electric part of the extinction torque (see, Appendix)
\begin{widetext}
\begin{equation}\label{Tmixe}
\av{\bs T_{\text{mix}}^{\,\text{e}}}=
\dfrac{\varepsilon_0}{2k^2}\,\text{Re}\! \displaystyle\oint_{4\pi}\!\!\!\td\Omega_u
\dfrac{e^{ikR}}{R}\!\!\displaystyle\oint_{S_\infty}\!\!\!
(\bs n\x\bs e^{\,*}_{\bs u}\,)\alpha_{\bs n}\,e^{-ikR(\bs u\,\.\,\bs n)} \,\td\sigma_n 
+\dfrac{\varepsilon_0}{2k}\,\text{Re}\! \displaystyle\oint_{4\pi}\!\!\!\td\Omega_u\,
e^{ikR}\!\!\displaystyle\oint_{S_\infty}\!\!\!
(\bs n\x\bs a_{\bs n})(\bs e^{\,*}_{\bs u}\.\bs n)
  \,e^{-ikR(\bs u\,\.\,\bs n)} \td\sigma_n,
\end{equation}
\end{widetext}
while the magnetic part can be deduced from ``the electric-magnetic democracy'' \cite{berry2009}.
The first term in $\av{\bs T_{\text{mix}}^{\,\text{e}}}$ can be evaluated by Jones' lemma
Eq.~(\ref{jones}), the second term requires an extension (see, Appendix), viz,
\begin{equation}\label{joneslin}
\begin{array}{ll}
\displaystyle\lim_{kR\to\infty}
\doint_{S_{\infty}}\bs G(\bs n) e^{-i k R(\bs u\.\bs n)}\,\td\sigma_n \\[4mm] \mspace{60mu}
\sim
\dfrac{\pi}{k^2} \bigl[\, e^{-ikR}\,\boldsymbol{\hat{L}}^2\bs G(\bs u) + e^{ikR}\,\boldsymbol{\hat{L}}^2\bs G(-\bs u)\,\bigr],
\end{array}
\end{equation}
for $\bs G(\bs u)=\bs G(-\bs u)=0$,
where
$\opL=-i\,\bs r\x\nabla$ is the orbital angular momentum operator (except for a factor $\hbar$)
\cite{schwinger,jackson,zangwill} acting on 
$\bs n$, and 
$$
\boldsymbol{\hat{L}}^2\bs G(\pm\bs u)\equiv \bigl[\boldsymbol{\hat{L}}^2\bs G(\bs n)\bigr]\biggl|_{\bs n=\pm \bs u}.
$$
The proof of Eq.~(\ref{joneslin}) follows the asymptotic behavior of integrals that can be evaluated by the method of stationary phase \cite{stamnes}.
It is given in (see, Appendix). 

The application of  the extended Jones' lemma Eq.~(\ref{joneslin}) to Eq.~(\ref{Tmixe}) needs 
explicit forms of the transverse and longitudinal far-field amplitudes
$\bs a_{\bs n}$ and $\alpha_{\bs n}$, which, together with $\bs b_{\bs n}$ and $\beta_{\bs n}$,
can be worked out based on the multipole expansion approach \cite{stratton,rose} and the irreducible tensor method \cite{silver}.
They are given by 
\begin{equation}\label{anbn}
\begin{array}{lll}
\bs a_{\bs n}=\dsum_{l=1}^{\infty}\left[\bs a_{\text{elec}}^{(l)}+\bs a_{\text{mag}}^{(l)}\right], & &
\alpha_{\bs n}=\dsum_{l=1}^{\infty}\alpha^{(l)}, \\[3mm]
\bs b_{\bs n}=\dsum_{l=1}^{\infty}\left[\bs b_{\text{elec}}^{(l)}+\bs b_{\text{mag}}^{(l)}\right], &\quad&
\beta_{\bs n}=\dsum_{l=1}^{\infty}\beta^{(l)}.
\end{array}
\end{equation}
where $\bs a_{\text{elec}}^{(l)}$ and $\bs a_{\text{mag}}^{(l)}$
[$\bs b_{\text{elec}}^{(l)}$ and $\bs b_{\text{mag}}^{(l)}$]
describe the transverse amplitudes of the scattered electric (magnetic) far-field
from an electric and a magnetic $2^l$-pole, respectively,
and $\alpha_{\bs n}$ ($\beta_{\bs n}$) represent the
longitudinal electric (magnetic) far-field due to an electric (a magnetic) $2^l$-pole.
The lower order cases with $l=1$, 2, 3, 4, and 5 correspond to the dipole,
quadrupole, octupole, hexadecapole, and dotriacontapole, respectively.
For a general order $l$, the scattered amplitudes read \revlin{(see, Appendix)} 
\begin{equation}\label{albl}
\begin{array}{llll}
\bs a_{\text{elec}}^{(l)}
&=& 
   i\,c\,\gamma_l\,\bs n\x \bigl(\bs n\x\bs{q}^{\,(l)}_{\text{elec}}\bigr), \mspace{20mu} \\[2mm]
\bs a_{\text{mag}}^{(l)}
&=& 
   i\,\gamma_l\,\bs n\x\bs{q}^{\,(l)}_{\text{mag}}, \\[2mm]
\bs b^{(l)}_{\text{elec}}
&=&\bs n\x \bs a^{(l)}_{\text{elec}},\quad\;\;
\bs b^{(l)}_{\text{mag}}
\,=\,\bs n\x \bs a^{(l)}_{\text{mag}}, \\[2mm]
\alpha^{(l)}
&=& 
-(l+1)\,c\,\gamma_l\,\bigl(\bs n\.\bs q^{\,(l)}_{\text{elec}}\bigr), \\[2mm]
\beta^{(l)}
&=& 
-(l+1)\,\gamma_l\,\bigl(\bs n\.\bs q^{\,(l)}_{\text{mag}}\bigr),
\end{array}
\end{equation}
with 
$
\gamma_l=\dfrac{(-ik)^{l+2}}{4\pi\varepsilon_0c\,l\,!}
$ and 
\begin{equation}
\begin{array}{lll}
\bs q^{\,(l)}_{\text{elec\,(mag)}}
=\bnn{(l-1)}\;\tc{l-1}\bOO{\text{elec\,(mag)}}{(l)}.
\end{array}
\label{qvec}
\end{equation}
Here $\bnn{(j)}$ denotes the $j$-fold tensor product of $\bs n$, yielding a rank-$j$ tensor.
The totally symmetric and traceless \cite{apple1989} rank-$l$ tensors
$\bOO{\text{elec}}{(l)}$ and $\bOO{\text{mag}}{(l)}$ are the electric and magnetic $2^l$-pole moments induced on the particle.
The relation between the 
multipole moments $\bOO{\text{elec(mag)}}{(l)}$
and the expansion coefficients of the scattered field in terms of
vector spherical wave functions \cite{stratton,jack2005} is exemplified in Appendix for some lower orders.
The multiple contraction between two tensors of ranks $l$ and $l'$, denoted by 
$\tc{m}$, results in a tensor of rank $l+l'-2m$ 
defined by 
\begin{widetext}
\begin{equation}\label{AtcB}
\begin{array}{ll}
\stackrel{\text{\scriptsize$\boldsymbol{\leftrightarrow}$}}
         {\boldsymbol{\mathbb{A}}}^{\text{\raisebox{-1.8ex}{$(l)$}}}
         \tc{m}
\stackrel{\text{\scriptsize$\boldsymbol{\leftrightarrow}$}}
         {\boldsymbol{\mathbb{B}}}^{\text{\raisebox{-1.8ex}{$(l')$}}}
=
\mathbb{A}^{(l)}_{i_1\,i_2\,\cdots\,i_{l-m}\,\textcolor{black}{{k_1\,k_2\,\cdots\,k_{m-1}\,k_m}}}\,
\mathbb{B}^{(l')}_{\textcolor{black}{{k_m\,k_{m-1}\,\cdots\,k_2\,k_1}}\,j_{m+1}\,\cdots\,j_{l'-1}\,j_{l'}}, \quad 0\le m\le\min\,[\,l,l'\,],
\end{array}
\end{equation}
where summation over repeated indices is assumed. 

With Eqs.~(\ref{joneslin}-\ref{qvec}), the integral over surface
$S_\infty$ in Eq.~(\ref{Tmixe}) can be taken (see, Appendix), leaving us with 
\begin{equation}
\av{\bs T^{\text{\,e}}_{\text{mix}}}
=\text{Re}\dsum_{l=1}^{\infty}
\dfrac{k^{l-1}}{2\, l!}\doint_{4\pi}\td\Omega_n(-i)^{l+1} 
  \bigl[(l-1)\bs n\x \bigl(\bqq{(l)}{\text{elec}}\.\bs e_{\bs n}^{\,*} \,\bigr)
         + \bs e_{\bs n}^{\,*}\x \bs q_{\text{elec}}^{\,(l)} \bigr]. 
\end{equation}
The magnetic part can be derived similarly, 
yielding the total extinction torque (see, Appendix)
\begin{eqnarray}
\av{\bs T_{\text{mix}}}
&\!\!=&\!\!
    \text{Re}\dsum_{l=1}^{\infty} \dfrac{1}{2\,l!}\bigl[(l-1)
 (\nabla^{(l-1)}\bs{E}_{\text{inc}}^*\,)\tc{l-1}\bOO{\text{elec}}{(l)}-
 \bOO{\text{elec}}{(l)}\tc{l-1}(\nabla^{(l-1)}\bs{E}_{\text{inc}}^*)\bigr]\tc{2}\beps\nonumber\\
&&\!\! {} +
  \text{Re}\dsum_{l=1}^{\infty}\dfrac{1}{2\,l!}\bigl[(l-1)
 (\nabla^{(l-1)}\bs{B}_{\text{inc}}^*\,)\tc{l-1}\bOO{\text{mag}}{(l)}
 -\bOO{\text{mag}}{(l)}\tc{l-1}(\nabla^{(l-1)}\bs{B}_{\text{inc}}^*)\bigr]\tc{2}\beps, \label{Tmix}
\end{eqnarray}
where 
$\beps$ is the Levi-Civita 
symbol 
\cite{zangwill}, and 
$\nabla^{(j)}\bs{V}$ denotes the $j$-fold gradient of vector $\bs V$. 

The recoil torque $\av{\bs T_{\text{sca}}}$ is similarly worked out to give (see, Appendix)
\begin{eqnarray}
\av{\bs T_{\text{sca}}}
&=&-\,\text{Im}\dsum_{l=1}^{\infty}\dfrac{k^{2l+1}}{8\pi\varepsilon_0} \,\dfrac{2^l(l+1)}{(2l+1)!}\;
 \bigl[\,
\bOO{\text{elec}}{(l)}\tc{l-1}\bOO{\text{elec}}{(l)*} 
+ \dfrac{1}{c^2}\,\bOO{\text{mag}}{(l)}\tc{l-1}\bOO{\text{mag}}{(l)*}\,\bigr]\tc{2}\beps. \label{Tsca}
\end{eqnarray}
\end{widetext}
The recoil troque 
turns out to depend solely on the contraction of 
the same multipole moment. 
This is contrary to
the recoil force. The latter arises from the 
coupling between different multipoles \cite{chen2011,chen2015}, 
either between electric (magnetic) multipoles of adjacent orders, or between electric and magnetic multipoles of the same order. 

The sum of $\av{\bs T_{\text{mix}}}$ and $\av{\bs T_{\text{sca}}}$ 
represents the multipole
expansion of the optical torque on any particle in generic monochromatic optical fields.
If one 
limits to the dipole terms, 
the torque reduces to (see, Appendix)
\begin{equation}\label{Tdip}
\mspace{-5mu}
\begin{array}{lll}
\av{\bs T_\text{dip}}
&\!\!=&\!\!\dfrac12\,\text{Re}(\bs p\x\bs E_{\text{inc}}^*)
   +\dfrac12\,\text{Re}(\bs m\x\bs B_{\text{inc}}^*)
    \\[2mm] &&
  \mspace{-25mu} {}
   +\dfrac{k^3}{12\pi\varepsilon_0}\,\text{Im}(\bs p\x\bs p^{\,*})
   +\dfrac{\mu_0k^3}{12\pi}\,\text{Im}(\bs m\x\bs m^*),
\end{array}\mspace{-5mu}
\end{equation}
where $\bs p=\bOO{\text{elec}}{(1)}$ [$\bs m=\bOO{\text{mag}}{(1)}$] is the electric (magnetic) dipole moment.
The recoil part, viz the last two terms in 
$\av{\bs T_\text{dip}}$, has been long missing in many previous studies
(e.g., \revlin{\cite{chaumet2009,ebbesen2013a,ebessen2013b,bliokh2014,bliokh2014a,beksh2015,bliokh2015}})
even in the dipolar limit.
Actually, it is the recoil torque that cancels out the extinction torque
on any non-absorbing spherical particle in arbitrary \revlin{harmonic} optical fields \cite{jiang2015,chen2014,antinieto},
not only limiting to plane wave fields \cite{marston}.

The right hand side of Eq.~(\ref{torque}) can be similarly written as a sum of the extinction and recoil terms, reading (see, Appendix)
\begin{equation}
-c\doint_{S_\infty} \av{\bs J_{\text{mix}}}\td\sigma -c\doint_{S_\infty} \av{\bs J_{\text{sca}}}  \td\sigma.  \label{Jab}
\end{equation}
By Eqs.~(\ref{eht}) and (\ref{ehinc}-\ref{ehs}),
the integrands can be expressed as functions of $\bs a_{\bs n}$, $\bs b_{\bs n}$, $\bs e_{\bs u}^{\,*}$ and $\bs e_{\bs u}^{\,*}$ (see, Appendix).
Using Jones lemma 
and its extension, as appropriate, and after lengthy algebra, one eventually arrives at (see, Appendix)
\begin{equation}
\begin{array}{ll}
-c\doint_{4\pi}\av{\bs J_{\text{mix}}}\td\sigma_n =\av{\bs T_{\text{mix}}}, \\[3mm]
-c\doint_{4\pi}\av{\bs J_{\text{sca}}}\td\sigma_n =\av{\bs T_{\text{sca}}},
\end{array}
\end{equation}
which finishes the proof of Eq. (\ref{torque}).

In a similar way, it can be proved that (see, Appendix)
\begin{equation}
\begin{array}{ll}
\av{\bs T_{\text{mix}}}
&=-\doint_{S}\bigl[\av{\bKKox}+\av{\bKKsx}\bigr]\.\bs n\td\sigma, \\[3mm]
\av{\bs T_{\text{sca}}}
&=-\doint_{S}\bigl[\av{\bKKos}+\av{\bKKss}\bigr]\.\bs n\td\sigma,
\end{array}
\end{equation}
and Eq.~(\ref{torque0}) follows from
$\doint_{S}\bigl[\av{\bKKoi}+\av{\bKKsi}\bigr]\.\bs n\td\sigma=0$.

To summarize,
just like the optical AM that is decomposed into two distinct parts
\cite{allen1992,allenbook2003,yao2011,humblet1943}: the orbital and the spin AM,
\revlin{the optical LM}
can be broken down into two parts of different natures
\revlin{as well} \cite{belinf1940,bliokh2014,bliokh2014a,beksh2015,bliokh2015,bliokh2016}:
the orbital (canonical)
momentum and the spin momentum, associated with the orbital and spin AM, respectively.
Based on  the method of stationary phase \cite{stamnes},
the multipole expansion approach \cite{stratton,rose}, and the irreducible tensor method \cite{silver},
we establish some universal relationships between the optical force/torque and the two contrasting types of optical LM/AM.
It is rigorously proved in generic case that
the optical force 
originated exclusively from
the transfer of \revlin{canonical momentum}
from light to matter, while
the spin momentum remains conserved even if the translational invariance is broken
by the introduction of particle into the optical fields,
in consistency with our understanding that the intrinsic nature of spin should not produce force.
Optical torque, on the other hand,
are brought about by the
total AM, namely, the sum of the orbital and spin AM components.
Additionally, we have presented, for the first time, a multipole expansion theory for optical torque
up to arbitrary order of multipoles.
Besides the extinction optical
torque originating from the interception of the incident photons as a simple extension to the static case, the theory
includes the long 
missing recoil torque that stems from
the interference of radiation from the multipoles of the same type and order induced on the particle in 
optical field.

\begin{acknowledgments}
The work is supported by the China 973 Projects (Grant No.
2013CB632701), NNSFC (No. 11574055). 
YKJ was supported by FDUROP.
\end{acknowledgments}


\appendix

\begin{widetext}

\noindent
\section{General monochromatic incident fields and the calculation of optical force and torque \label{appa}} 

In this section we demonstrate that Eqs.~(9) in the main text represent a generic monochromatic optical field in free-space for light scattering problem.

In free-space, any monochromatic incident electric and magnetic fields can be expanded in terms of vector spherical wave functions (VSWFs) 
\begin{equation} \label{EHinc0}
\begin{array}{l}
\bs E_{\text{inc}}=-\dsum_{l=1}^{\infty}\dsum_{m=-l}^{l} i^{l+1}C_{ml}E_0
      \Bigl[\,p_{m,l}\bN_{ml}^{(1)}(k,\br)
            + q_{m,l}\bM_{ml}^{(1)}(k,\br)\,\Bigr]  \\[3mm]
\bs H_{\text{inc}}=-\dfrac{1}{Z_0} \dsum_{l=1}^{\infty}\dsum_{m=-l}^{l}  i^{l}C_{ml}E_0
      \Bigl[\,q_{m,l}\bN_{ml}^{(1)}(k,\br)
           + p_{m,l}\bM_{ml}^{(1)}(k,\br)\,\Bigr],
\end{array}
\end{equation}
since both fields are divergence-free \cite{stratton}.
Here $Z_0=\sqrt{\mu_0/\varepsilon_0}$ is the wave impedance in free-space, $E_0$ characterizes the incident field amplitude,
and
\begin{equation}
C_{ml}=\left[\frac{(2l+1)}{l(l+1)} \frac{(l-m)!}{(l+m)!}\right]^{1/2}.
\end{equation}
So arbitrary monochromatic optical fields are completely characterized by
the coefficients $p_{m,l}$ and $q_{m,l}$, which
are known as the partial wave expansion coefficients, or beam shape coefficients in the generalized Lorenz-Mie theory \cite{glmt}.

The VSWFs $\bM^{(1)}_{ml}$ and $\bN^{(1)}_{ml}$ in Eqs.~(\ref{EHinc0}) are regular ones, 
given by, see, {\it e.g.}, \cite{bohren1998,tsang,wittmann},
\begin{equation}\label{VSWF1}
\begin{array}{ll}
\bM^{(1)}_{ml}(k,\br)&\!\!=
      \bigl[\,i\pi_{ml}(\cos\theta)\etheta-\tau_{ml}(\cos\theta)\ephi\,\bigr]
        \dfrac{\psi_l(kr)}{kr}e^{im\phi},  \\[3mm]
\bN^{(1)}_{ml}(k,\br) &\!\!=
      \bigl[\,\tau_{ml}(\cos\theta)\etheta+i\pi_{ml}(\cos\theta)\ephi\,\bigr]
          \dfrac{\psi'_l(kr)}{kr}e^{im\phi}  + \er l(l+1) P^m_l(\cos\theta) \dfrac{\psi_l(kr)}{(kr)^2}e^{im\phi},
\end{array}
\end{equation}
where the two auxiliary functions, $\pi_{ml}(\cos\theta)$ and
$\tau_{ml}(\cos\theta)$, are defined by
\begin{equation}
\pi_{ml}(\cos\theta)=\frac{m}{\sin\theta} P^m_l(\cos\theta), \quad
\tau_{ml}(\cos\theta)=\frac{\td}{\td\theta} P^m_l(\cos\theta),
\end{equation}
with $P^m_l(x)$ denoting the associated Legendre function of the
first kind and $\psi_l(x)$ the Riccati Bessel function \cite{bohren1998}.

The regular VSWFs 
$\bM^{(1)}_{ml}$ and $\bN^{(1)}_{ml}$ 
can be written as integral representation of vector spherical harmonics \cite{stratton,tsang,wittmann}
\begin{equation} \label{vswf2vplw}
\begin{array}{ll}
\bM^{(1)}_{ml}(k,\br)&\!\!=\dfrac{(-i)^{l+1}}{\sqrt{4\pi}\,C_{m,l}}\doint_{4\pi}\bX_{lm}(\bs u)e^{i k \bs u\.\bs r}\td\Omega_u
\\[3.5mm]
\bN^{(1)}_{ml}(k,\br) &\!\!=\dfrac{(-i)^{l}}{\sqrt{4\pi}\,C_{m,l}}\doint_{4\pi}\bs u\x\bX_{lm}(\bs u)e^{i k \bs u\.\bs r}\td\Omega_u
\end{array}
\end{equation}
where the integration is over the entire solid angle of $4\pi$ for the directions of wave vectors $\bs k=k \bs u$,
with $\bs u$ being the real unit vector and $k=\omega \sqrt{\varepsilon_0\mu_0}$ the wave number.
The vector spherical harmonics $\bX_{lm}(\bs u)$ is defined by \cite{tsang,wittmann,jackson}
$$
\bX_{lm}(\theta,\phi)=\dfrac1{\sqrt{l(l+1)}}\,\opL Y_{lm}(\theta,\phi)
$$
where $\opL=-i\,\bs r\x\nabla$ is the orbital angular momentum operator and $Y_{lm}(\theta,\phi)$ denotes the spherical harmonics \cite{jackson}. 

Inserting Eqs. (\ref{vswf2vplw}) into the incident fields (\ref{EHinc0}) yields
\begin{equation} \label{EHinc1}
\begin{array}{l}
\bs E_{\text{inc}}=\doint_{4\pi}\biggl\{-\dfrac{E_0}{\sqrt{4\pi}}\dsum_{l=1}^{\infty}\dsum_{m=-l}^{l}
      \Bigl[i\,p_{m,l}\, \bs u\x\bX_{lm}(\bs u)
            + q_{m,l}\bX_{lm}(\bs u)\,\Bigr]  \biggr\}\,e^{ik\bs u\.\bs r}\td\Omega_u
            =\doint_{4\pi} \be_{\bs u} \,e^{i k \bs u\.\bs r}\,\td\Omega_u
             \\[3mm]
\bs H_{\text{inc}}=\dfrac{1}{Z_0}\,\doint_{4\pi}\biggl\{-\dfrac{E_0}{\sqrt{4\pi}} \dsum_{l=1}^{\infty}\dsum_{m=-l}^{l}
      \Bigl[\,q_{m,l}\,\bs u\x\bX_{lm}(\bs u)
           -i\, p_{m,l}\bX_{lm}(\bs u)\,\Bigr]\biggr\}\,e^{ik\bs u\.\bs r}\td\Omega_u
           =\dfrac{1}{Z_0}\doint_{4\pi} \bs h_{\bs u} \,e^{i k \bs u\.\bs r}\,\td\Omega_u
\end{array}
\end{equation}
which is of the form of Eqs.~(9) in the main text.
Here $\be_{\bs u}$ and $\bh_{\bs u}$ denote, respectively, the
parts within curly braces in the integrands for
$\bs E_{\text{inc}}$ and $\bs H_{\text{inc}}$. They satisfy
$\bs u\x\be_{\bs u}=\bs h_{\bs u}$ and  $\bh_{\bs u}\x\bs u=\bs e_{\bs u}$,

Some remarks concerning the calculation of optical force and torque 
are in order.

Equations (\ref{EHinc0}) and, equivalently, Eqs.~(\ref{EHinc1}),
describe arbitrary optical field in the sourceless region where a particle of arbitrary shape and size is to be put in. 
Based on the generalized Lorenz-Mie theory \cite{glmt} and the T-matrix method \cite{Tmatrix},
one can then solve for the fields $\bs E_s$ and $\bs H_s$ that are
scattered off the particle when the latter is introduced. 
The results read, in terms of VSWFs,
\begin{equation} \label{EHsct0}
\begin{array}{l}
\bs E_s=\dsum_{l=1}^{\infty}\dsum_{m=-l}^{l} i^{l+1}C_{ml}E_0
      \Bigl[\,a_{m,l}\bN_{ml}^{(3)}(k,\br)
            + b_{m,l}\bM_{ml}^{(3)}(k,\br)\,\Bigr],  \\[3mm]
\bs H_s=\dfrac{1}{Z_0} \dsum_{l=1}^{\infty}\dsum_{m=-l}^{l}  i^{l}C_{ml}E_0
      \Bigl[\,b_{m,l}\bN_{ml}^{(3)}(k,\br)
           + a_{m,l}\bM_{ml}^{(3)}(k,\br)\,\Bigr],
\end{array}
\end{equation}
where the outgoing VSWFs $\bM^{(3)}_{ml}$ and $\bN^{(3)}_{ml}$, describing the multipole fields \cite{stratton,rose},
are given by (\ref{VSWF1}) with the replacement of the Riccati Bessel functions $\psi(x)$
by the Riccati Hankel functions $\xi(x)$ \cite{bohren1998}.
The expansion coefficients $a_{m,l}$ and $b_{m,l}$ in Eqs. (\ref{EHsct0}) are related to the beam shape coefficients
$p_{m,l}$ and $q_{m,l}$ in (\ref{EHinc0}) by the T-matrix \cite{Tmatrix}.

Next, the optical force and torque are evaluated, respectively,
by the surface integrals of the linear momentum (LM) and angular momentum (AM) current density tensor
over the outer surface of the particle.
Both the tensors are evaluated based on the true optical fields $\bs E_t=\bs E_{\text{inc}}+\bs E_s$ and $\bs H_t=\bs H_{\text{inc}}+\bs H_s$ immediately outside the particle,
which are determined by the superposition of the incident fields Eqs.~(\ref{EHinc0}) and the scattered fields Eqs.~(\ref{EHsct0}).
As both LM and AM current density tensors so obtained are divergence-free exterior to the particle,
a consequence following, physically, from the conservation of LM and AM in free space 
and guaranteed, mathematically, by the properties of VSWFs,
the surface integrals can be performed over a spherical surface $S_{\infty}$ with radius $R\to\infty$ centered at
the particle, while, most importantly, keeping the fields still expressed as a superposition of
those given by Eqs.~(\ref{EHinc0}) and Eqs.~(\ref{EHsct0}).
It is noted that at the surface $S_{\infty}$, 
the true physical fields are actually not always 
given by the pseudo-fields $\bs E_t$ and $\bs H_t$ there,
but it is these pseudo-fields that one should use to perform the surface integrals if
the optical force and torque on the specific particle are to be calculated.
If we evaluate LM and AM current density tensors based on the true physical fields
at $S_{\infty}$, we are left with the total optical force and torque that exert on all scatterers, including light source, enclosed
by $S_\infty$, instead of those on the specific particle under study.
As a result, for the purpose of attacking 
the light scattering problem, a generic monochromatic optical fields
can always be described by (\ref{EHinc0}), and, equivalently, by (\ref{EHinc1}), provided that
the influence due to the multiple scattering between the introduced particle and other particles (including light sources)
can be neglected. In most cases of optical micromanipulation, the influence of the multiple scattering is negligible.
If the effect of the multiple scattering is significant, the equivalent incident fields,
which sum all fields originating from both the light sources and all other particles,
should be worked out self-consistently based on the multiple scattering theory, see, \textit{e.g.}, Refs.~\cite{martin,mishchenko2006,mishchenko2014}.
In the latter case, the equivalent incident fields on a specific particle can still be cast into the form of (\ref{EHinc1}). This indeed constitutes
the basic idea of the multiple scattering theory \cite{martin,mishchenko2006,mishchenko2014}.

Finally, it is noted that, despite appearing in a form reminiscent of the angular spectrum representation that
excludes evanescent wave components \cite{mandel}, 
 Eqs.~(\ref{EHinc1}) are not simply
a representation of plane wave spectrum for optical fields, but
possess more general physical significance in the light scattering theory. They actually
denote a generic monochromatic optical field,
with the evanescent wave components already taken into account.

\vskip 20pt

\noindent
\section{Jones lemma and its extension} 

In this section we present a proof for the extended Jones lemma Eq.~(19) in the main text.

The asymptotic behavior of the double integral $J$, with
\begin{equation} \label{Jdbl}
J=\dint\!\!\!\dint_D g(x,y) e^{iKf(x,y)}\td x\td y,
\end{equation}
is given, when positive $K\to\infty$, by \cite{stamnes}
\begin{equation}\label{Js}
J=\dsum_sJ_s, \quad \text{with} \quad J_s\sim \dfrac{2\pi\sigma}{K|H|^{1/2}}\,e^{iK f(x_s,y_s)}\left(Q_0+\frac iK Q_2\right),
\end{equation}
where the summation runs over the interior stationary point $(x_s,y_s)$ at which the partial
derivatives $f_x$ and $f_y$ of the phase function $f(x,y)$ vanishes, \textit{i.e.},
$f_x(x_s,y_s)=f_y(x_s,y_s)=0$, whereas \cite{stamnes}
\begin{subequations} \label{stapnt}
\begin{eqnarray}
\sigma
&=&\exp\,\Bigl\{\,i\,\dfrac\pi4\bigl[\text{sgn}\,(F_{2,0})+\text{sgn}\,(F_{0,2})\bigr]\,\Bigr\}
=\left\{ \begin{array}{lll}
1  && \text{if $H<0$} \\
i  && \text{if $H>0$ and $f_{0,2}>0$} \\
-i && \text{if $H>0$ and $f_{0,2}<0$}
\end{array} \right.       \label{stapnta}              \\[3mm]
Q_0&=&G_{0,0}=g_{0,0},\qquad\qquad H=4 F_{2,0}F_{0,2}=4f_{2,0}f_{0,2}-f_{1,1}^2   \label{stapntb} 
\end{eqnarray}
\begin{eqnarray}
Q_2&=&
\dfrac{4G_{0,0}}{H^3}\biggl\{
       15(F_{2,0}^3F_{0,3}^2+F_{0,2}^3F_{3,0}^2)+
       \underline{\frac14H}\Bigl[3F_{0,2}(2F_{1,2}F_{3,0}+F_{2,1}^2) \nonumber \\[1mm]
& & \begin{array}{lll}
\mspace{50mu} {}+\underline{3F_{2,0}}(2F_{2,1}F_{0,3}+F_{1,2}^2)-
12(F_{2,0}^2F_{0,4}+F_{0,2}^2F_{4,0})-4F_{2,0}F_{0,2}F_{2,2}\Bigr]\biggr\}\\[3mm]
 \mspace{-5mu} {}-\dfrac{4}{H^2}\Bigl[G_{1,0}(F_{2,0}F_{0,2}F_{1,2}+3F_{0,2}^2F_{3,0})
                                      + G_{0,1}(F_{0,2}F_{2,0}F_{2,1}+3F_{2,0}^2F_{0,3})\Bigr] 
                                      +\dfrac{2}{H}(G_{0,2}F_{2,0}+ G_{2,0}F_{0,2} ),
\end{array} \label{stapntc}
\end{eqnarray}
with
\begin{equation} \label{stapnte}
\mspace{-5mu}
\begin{array}{l}
G_{0,0}=g_{0,0}, \quad G_{1,0}=g_{1,0}+\tilde{c}g_{0,1}, \quad
G_{0,1}=g_{0,1},\quad G_{2,0}=g_{2,0}+\tilde{c}g_{1,1}+\tilde{c}^2g_{0,2}, \quad G_{0,2}=g_{0,2}, \\[2mm]
F_{0,0}=f_{0,0}, \quad F_{2,0}=f_{2,0}-\tilde{c}^2f_{0,2}, \quad F_{0,2}=f_{0,2}, \quad
F_{3,0}=f_{3,0}+\tilde{c}f_{2,1}+\tilde{c}^2f_{1,2}+\tilde{c}^3f_{0,3} , \\[2mm]
F_{2,1}=f_{2,1}+2\tilde{c}f_{1,2}+3\tilde{c}^2f_{0,3}, \qquad\;\; F_{1,2}=f_{1,2}+3\tilde{c}f_{0,3}, \qquad \quad \;\; F_{0,3}=f_{0,3}, \\[2mm]
F_{4,0}=f_{4,0}+\tilde{c}f_{3,1}+\tilde{c}^2f_{2,2}+\tilde{c}^3f_{1,3}+\tilde{c}^4f_{0,4}, \qquad\;\;
F_{3,1}=f_{3,1}+2\tilde{c}f_{2,2}+3\tilde{c}^2f_{1,3}+4\tilde{c}^3f_{0,4}, \\[2mm]
F_{2,2}=f_{2,2}+3\tilde{c}f_{1,3}+6\tilde{c}^2f_{0,4}, \qquad\;\; F_{1,3}=f_{1,3}+4\tilde{c}f_{0,4}, \qquad \quad \;\; F_{0,4}=f_{0,4}, \\[3mm]
\tilde{c}=-\dfrac{f_{1,1}}{2f_{0,2}}\biggl|_{x_s,\,y_s}\mspace{-5mu}, \qquad
f_{m,n}=\dfrac{1}{m!\,n!}\,\dfrac{\partial^{\,m+n}f(x,y)}{\partial x^m\, \partial y^n}\biggl|_{x_s,\,y_s}\mspace{-5mu}, \qquad
g_{m,n}=\dfrac{1}{m!\,n!}\,\dfrac{\partial^{\,m+n}g(x,y)}{\partial x^m\, \partial y^n}\biggl|_{x_s,\,y_s}\mspace{-5mu}.
\end{array}\mspace{-7mu}
\end{equation}
\end{subequations}

\vskip- 5pt
\noindent
Here $\tilde{c}$ is introduced to get rid of a cross term \cite{stamnes,dingle}.
Some misprints in \cite{stamnes} have been corrected here in Eq.~(\ref{stapntc}), marked by underline.
For the present case, keeping only the $Q_0$ term in Eq.~(\ref{Js})
yields Jones lemma Eq.~(10) in the main text (see, {\it e.g.}, Appendix XII of \cite{born}), viz 
\begin{equation} \label{jonesapp}
\begin{array}{ll}
\displaystyle\lim_{kR\to\infty}\frac1{kR}\oint_{S_{\infty}}
\bs G(\bs n) e^{-i k (\bs u\.\bs n)R}\,\td\sigma
\sim
\dfrac{2\pi i}{k^2} \bigl[\, \bs G(\bs u)e^{-ikR} - \bs G(-\bs u)e^{ikR}\,\bigr],
\end{array}
\end{equation}
where $\bs G(\bs n)$ is an arbitrary function of $\bs n$,
whereas $\bs u$ is an arbitrary real unit vector, and
$\bs n$ is a unit vector in the local outward radial direction
of the spherical surface $S_\infty$ with large radius $R\to\infty$.
To evaluate the second term in $\av{\bs T_{\text{mix}}^{\text{\,e}}}$,
given in Eq. (16) of the main text and recapitulated below,
\begin{equation}\label{Tmixeapp}
\dfrac{\varepsilon_0}{2k}\,\text{Re}\! \displaystyle\oint_{4\pi}\!\!\!\td\Omega_u\,
e^{ikR}\!\!\displaystyle\oint_{S_\infty}\!\!\!
(\bs n\x\bs a_{\bs n})(\bs e^{\,*}_{\bs u}\.\bs n)
  \,e^{-ikR\,(\bs u\,\.\,\bs n)} \td\sigma_n,
\end{equation}
one needs to evaluate $Q_2$ in Eq.~(\ref{Js}).

In the spherical polar coordinate system, by writing the Cartesian components of the unit vectors $\bs u$ and $\bs n$ in the form
\begin{equation}
\bs u=(\sin\theta_0\cos\phi_0,\, \sin\theta_0\sin\phi_0, \,\cos\theta_0),\qquad
\bs n=(\sin\theta\cos\phi,\, \sin\theta\sin\phi,\, \cos\theta),
\end{equation}
with $\theta$ and $\phi$ denoting, respectively, the polar and azimuthal angles,
the integral over $S_\infty$ in Eq.~(\ref{Tmixeapp}) can be cast into the form given in Eq.~(\ref{Jdbl}) by
\begin{subequations} \label{Lb}
\begin{equation}
\displaystyle\oint_{S_\infty}\bs G(\bs n)e^{-i k R(\bs u\.\bs n)}\td\sigma_n
=\dint_{\theta=0}^{\pi}\dint_{\phi=0}^{2\pi}\! g(\theta,\phi)e^{iKf(\theta,\phi)}\,\td\phi\td \theta,
\end{equation}
with $K=kR$, \ $g(\theta,\phi)=R^2\bs G(\bs n) \sin\theta$, and
\begin{equation}
f(\theta,\phi)=-\bs u\.\bs n=-\bigl[\cos\theta\cos\theta_0+\sin\theta\sin\theta_0\cos(\phi-\phi_0)\bigr].
\end{equation}
\end{subequations}
The stationary points are thus determined by
\begin{equation}
\pd{f(\theta,\phi)}{\theta}=\pd{f(\theta,\phi)}{\phi}=0,
\end{equation}
which yields
\begin{equation}
\left\{\begin{array}{l}
(\theta_1,\phi_1)=(\theta_0,\phi_0),\\[2mm]
(\theta_2,\phi_2)=(\pi-\theta_0,\pi+\phi_0),
\end{array}\right. \quad \text{corresponding, respectively, to} \quad
\left\{
\begin{array}{l}
\bs n_1=\bs u,\\[2mm]
\bs n_2=-\bs u.
\end{array}\right.
\end{equation}
Since at either stationary point,
$f_{1,1}=\dfrac{\partial^2f(\theta,\phi)}{\partial\theta\,\partial\phi}\biggl|_{\theta_i,\phi_i}=0$, for $i=1,2,$
one has $\tilde{c}=0$ and Eq.~(\ref{stapnt}) can be much simplified. When $\bs G(\bs u)=0$ and $\bs G(-\bs u)=0$, like
the integral over $S_\infty$ in Eq.~(\ref{Tmixeapp}), the leading term $Q_0$ vanishes.
The integral in Eq.~(\ref{Lb}a) then reduces, after some straightforward algebra based on Eqs.~(\ref{stapnt}), to
\begin{equation}\label{joneslinapp}
\displaystyle\lim_{kR\to\infty}\displaystyle\oint_{S_\infty}\bs G(\bs n)e^{-i k R(\bs u\.\bs n)}\td\sigma_n
\sim
\dfrac{\pi}{k^2} \bigl[\, e^{-ikR}\,\boldsymbol{\hat{L}}^2\bs G(\bs u) + e^{ikR}\,\boldsymbol{\hat{L}}^2\bs G(-\bs u)\,\bigr],
\end{equation}
where $\opL$ is the orbital angular momentum operator,
\begin{equation}
\opL^2=-\dfrac1{\sin\theta}\pd{}{\theta}\left(\sin\theta\pd{}{\theta}\right)-\dfrac1{\sin^2\theta}\,\pd{^2}{\phi^2},
\end{equation}
and we have used the notation
$$
\opL^2\bs G(\pm\bs u)=\bigl[\boldsymbol{\hat{L}}^2\bs G(\bs n)\bigr]\biggl|_{\bs n=\pm \bs u}.
$$
This concludes the proof of the extended Jones' lemma.
It is interesting 
to note that the extended Jones lemma, which will be used in
the calculation of the optical torque, involves
the angular momentum operator $\boldsymbol{\hat{L}}$.

\vskip 20pt

\noindent
\section{Optical force} 

In this section, we derive the expressions for the extinction and recoil forces, the surface integrals of
orbital momentum $\av{\bs M^o_{\text{mix}}}$ and $\av{\bs M^o_{\text{sca}}}$, and the surface integrals of
orbital momentum current density $\av{\bTTox}$ and  $\av{\bTTos}$,
namely,  Eqs. (12), (14), and (16) in the main text.

Based on the Maxwell stress tensor formalism \cite{jackson,zangwill,schwinger} (see, also, discussion in Appendix \ref{appa}),
a decomposition of the total fields into the incident and scattered fields,
\begin{equation} \label{ehtapp}
\bs E_{\text{tot}} =\bs E_{\text{sca}} +\bs E_{\text{inc}} \quad \text{and}\quad
\bs H_{\text{tot}} =\bs H_{\text{sca}} +\bs H_{\text{inc}},
\end{equation}
suggests that
\begin{equation}
\begin{array}{lll}
\av{\bs F}=\av {\bs F_\text{inc}}+\av {\bs F_\text{mix}}+\av {\bs F_\text{sca}}, & \quad \qquad &
\av {\bs F_\text{inc}}
 =\displaystyle\oint_{S_\infty}\!\!\bs n\.\av{\bTT_\text{inc}}\td\sigma, \\[2mm]
 \av {\bs F_{\text{mix}}}
 =\displaystyle\oint_{S_\infty}\!\!\bs n\.\av{\bTT_{\text{mix}} }\td\sigma, &&
 \av {\bs F_\text{sca}}
 =\displaystyle\oint_{S_\infty}\!\!\bs n\.\av{\bTT_\text{sca} }\td\sigma
\end{array}
\end{equation}
where $\av{\bTT_\text{inc}}$ involves the incident fields only,
$\av{\bTT_\text{sca}}$ depends solely on the scattered fields, while
$\av{\bTT_{\text{mix}}}$ includes all the rest (mixed) terms. To be specific,
with the use of Re$[\bs E_{\text{inc}}\bs E_{\text{sca}}^*]=\text{Re}[\bs E_{\text{inc}}^*\bs E_{\text{sca}}]$ etc., they are given by
\begin{equation}\label{TiTmTs}
\mspace{-5mu}
\begin{array}{lll}
\av{\bTT_\text{inc} }
&= \dfrac12\,\text{Re}
\bigl[\varepsilon_0\bs E_{\text{inc}}\bs E_{\text{inc}}^*
+\mu_0\bs H_{\text{inc}}\bs H_{\text{inc}}^* 
-\dfrac12\Bigl(\varepsilon_0\bs E_{\text{inc}}\.\bs E_{\text{inc}}^*
 +\mu_0\bs H_{\text{inc}}\.\bs H_{\text{inc}}^*\Bigr)\bI\,
\bigr], \\[3mm]
 \av{\bTT_\text{mix} }
 &=\dfrac12\,\text{Re}
\bigl[(
\varepsilon_0\bs E_{\text{inc}}^*\bs E_{\text{sca}}+
\varepsilon_0\bs E_{\text{sca}}\bs E_{\text{inc}}^*+
\mu_0\bs H_{\text{inc}}^*\bs H_{\text{sca}}+
\mu_0\bs H_{\text{sca}}\bs H_{\text{inc}}^*) 
{}-(\varepsilon_0\bs E_{\text{inc}}^*\.\bs E_{\text{sca}}+\mu_0\bs H_{\text{inc}}^*\.\bs H_{\text{sca}})\bI\,
\bigr], \\[3mm]
 \av{\bTT_\text{sca} }
 &=\dfrac12\,\text{Re}
\bigl[
\varepsilon_0 \bs E_{\text{sca}}\bs E_{\text{sca}}^*+
\mu_0 \bs H_{\text{sca}}\bs H_{\text{sca}}^*-\dfrac12
(\varepsilon_0 \bs E_{\text{sca}}\.\bs E_{\text{sca}}^*+
 \mu_0          \bs H_{\text{sca}}\.\bs H_{\text{sca}}^*)\bI\,\bigr]\td\sigma,
\end{array}\mspace{-5mu}
\end{equation}
The incident part
$\av{\bTT_\text{inc}}$ gives no net contribution to the time-averaged optical force, as can be understood
by the momentum conservation law for monochromatic fields in non-absorptive space
(see, e.g., \S13.3 and Appendix XI of \cite{born} for a similar discussion on energy flux), viz,
$$
\begin{array}{lll}
\av {\bs F_\text{inc}}
&\!\!= \displaystyle\oint_{S_\infty}\!\!\bs n\.\av{\bTT_\text{inc} }\td\sigma=0. 
\end{array}
$$
The extinction force \cite{bohren} is
\begin{equation}
\begin{array}{lll}
\av {\bs F_{\text{mix}}}
&\!\!= \dfrac12\,\text{Re}\displaystyle\oint_{S_\infty}
\Bigl[\bs n\.(
\varepsilon_0\bs E_{\text{inc}}^*\bs E_{\text{sca}}+
\varepsilon_0\bs E_{\text{sca}}\bs E_{\text{inc}}^*+
\mu_0\bs H_{\text{inc}}^*\bs H_{\text{sca}}+
\mu_0\bs H_{\text{sca}}\bs H_{\text{inc}}^*) \\ & \mspace{95mu}
{}-(\varepsilon_0\bs E_{\text{inc}}^*\.\bs E_{\text{sca}}+\mu_0\bs H_{\text{inc}}^*\.\bs H_{\text{sca}})\,\bs n\,
\Bigr]\td\sigma.
\end{array}
\end{equation}
Inserting Eq. (\ref{EHinc1})
for monochromatic incident fields,
\begin{equation} \label{ehincapp}
\bs E_{\text{inc}}=\oint_{4\pi} \be_{\bs u} \,e^{i k \bs u\.\bs r}\,\td\Omega_u,  \qquad 
\bs H_{\text{inc}}=\dfrac1{Z_0}\oint_{4\pi} \bh_{\bs u} \,e^{i k \bs u\.\bs r}\,\td\Omega_u,
\end{equation}
and the expressions for the scattered fields at large distance
from any particle (see, e.g., \S13.6 of \cite{born}) 
\begin{equation} \label{ehsapp}
\bs E_{\text{sca}}= \left(\bs a_{\bs n}+\dfrac{\bs n\,\alpha_{\bs n}}{kr}\right)\dfrac{e^{ikr}}{kr}, \qquad 
\bs H_{\text{sca}}= \dfrac1{Z_0}\left(\bs b_{\bs n}+\dfrac{\bs n\,\beta_{\bs n}}{kr}\right) \dfrac{e^{ikr}}{kr},
\end{equation}
it is not difficult to obtain the integral of the first two terms in the integrand of $\av {\bs F_{\text{mix}}}$
$$
\begin{array}{ll}
\bs f_1
&=\dfrac{\varepsilon_0}2\,\text{Re}\displaystyle\oint_{S_\infty}
\bigl[
\bs n\.\bs E_{\text{inc}}^*\bs E_{\text{sca}}+
\bs n\.\bs E_{\text{sca}}\bs E_{\text{inc}}^*]\td\sigma\\[2mm]
&=
\dfrac{\varepsilon_0}2 \,\text{Re}  \displaystyle\oint_{4\pi}\td\Omega_u
\displaystyle\oint_{S_\infty}\!\!
\bigl[
(\bs e^{\,*}_{\bs u}\.\bs n) \bs a_{\bs n}+
(\bs a_{\bs n}\.\bs n) \bs e^{\,*}_{\bs u}\,\bigr]
\dfrac{e^{ikr}}{kr}\,e^{-ik(\bs u\.\bs r)} \,\td\sigma   \\[2mm]
&=
\dfrac{\pi\varepsilon_0}{k^2}\, \text{Re} \displaystyle\oint_{4\pi}i
\bigl[
(\bs e_{\bs u}^{\,*}\.\bs u)\bs a_{\bs u}+(\bs e_{\bs u}^{\,*}\.\bs u)\bs a_{-\bs u}\,e^{2ikR}
\bigr]\td\Omega_u=0,
\end{array}
$$
where use has been made of Jones lemma Eq.~(\ref{jonesapp}) for the surface integral over $S_\infty$ and, also, some of the following relations, as appropriate
\begin{equation}\label{anbneuhu}
\bs u\.\bs e_{\bs u}=\bs u\.\bs h_{\bs u}=0,\;\;
\bh_{\bs u} = \bs u\x\bs e_{\bs u},\quad
\bs n\.\bs a_{\bs n}=\bs n\.\bs b_{\bs n}=0,\;\;
\bs b_{\bs n}=\bs n \x \bs a_{\bs n}.
\end{equation}
Similarly, the integral of the latter two terms in the integrand of $\av {\bs F_{\text{mix}}}$ also vanishes,
$$
\begin{array}{ll}
\bs f_2
&=\dfrac{\mu_0}2\,\text{Re}\displaystyle\oint_{S_\infty}
\bigl[
\bs n\.\bs H_{\text{inc}}^*\bs H_{\text{sca}}+
\bs n\.\bs H_{\text{sca}}\bs H_{\text{inc}}^*\bigr]\td\sigma=0. 
\end{array}
$$
The integral of the last two terms in $\av{\bs F_{\text{mix}}}$ reduces to
$$\begin{array}{ll}
\bs f_3
&=-\dfrac{1}2\,\text{Re}\displaystyle\oint_{S_\infty}
\bigl[
\varepsilon_0\bs E_{\text{inc}}^*\.\bs E_{\text{sca}}+
\mu_0\bs H_{\text{inc}}^*\.\bs H_{\text{sca}}\bigr]\,\bs n\,\td\sigma \\[3mm]
&=-
\dfrac{\varepsilon_0}2 \,\text{Re} \displaystyle\oint_{4\pi}\td\Omega_u
\,\dfrac{e^{ikR}}{kR}\displaystyle\oint_{S_\infty}\!\!
\bigl[
(\bs e^{\,*}_{\bs u}\. \bs a_{\bs n})\,\bs n\, e^{-ikR\,(\bs u\.\bs n)}+
(\bs h^{\,*}_{\bs u}\. \bs b_{\bs n})\,\bs n\, e^{-ikR\,(\bs u\.\bs n)}\bigr]
 \,\td\sigma  \\[3mm]
&=-\dfrac{\pi\varepsilon_0}{k^2}\, \text{Re}
 \displaystyle\oint_{4\pi} i
\bigl[
 (\bs e_{\bs u}^{\,*}\.\bs a_{ \bs u})
+(\bs e_{\bs u}^{\,*}\.\bs a_{-\bs u}) e^{2ikR}
+(\bs h_{\bs u}^{\,*}\.\bs b_{ \bs u})
+(\bs h_{\bs u}^{\,*}\.\bs b_{-\bs u}) e^{2ikR}
\bigr]\bs u \,\td\Omega_u\\[3mm]
&=-\dfrac{\pi\varepsilon_0}{k^2}\, \text{Re}
 \displaystyle\oint_{4\pi} i
\bigl[
 (\bs e_{\bs u}^{\,*}\.\bs a_{ \bs u})
+(\bs e_{\bs u}^{\,*}\.\bs a_{-\bs u}) e^{2ikR}
+      (\bs e^{\,*}_{\bs u}\.\bs a_{ \bs u})
-      (\bs e^{\,*}_{\bs u}\.\bs a_{-\bs u}) e^{2ikR}
\bigr]\bs u \,\td\Omega_u \\[3mm] &
=\dfrac{2\pi\varepsilon_0}{k^2}\,\text{Im}
\displaystyle\oint_{4\pi}
(\be^{\,*}_{\bs u}\.\bs a_{\bs u})\bs u\,\td\Omega_u,
\end{array}
$$
where one has used once again Jones lemma Eq.~(\ref{jonesapp}) for the surface integral over $S_\infty$ and, also,
$\bs h^{\,*}_{\bs u}\.\bs b_{\pm\bs u}=\pm \bs e^{\,*}_{\bs u}\.\bs a_{\pm\bs u}$.
Putting together, one has the first equation of Eq.~(12) in the main text, viz
\begin{equation}\label{FTFS}
\av {\bs F_{\text{mix}}}=\bs f_1+\bs f_2+\bs f_3=
\dfrac{2\pi\varepsilon_0}{k^2}\,\text{Im}\displaystyle\oint_{4\pi}
(\be^{\,*}_{\bs u}\.\bs a_{\bs u})\bs u\,\td\Omega_u.
\end{equation}
The recoil force reads, with the use of  $|\bs a_{\bs n}|^2=|\bs b_{\bs n}|^2$ that follows Eq.~(\ref{anbneuhu}),
\begin{equation}
\begin{array}{lll}
\av {\bs F_\text{sca}}
&= \dfrac12\,\text{Re}\displaystyle\oint_{S_\infty}
\Bigl[
\varepsilon_0\bs n\.\bs E_s\bs E_s^*+
\mu_0\bs n\.\bs H_s\bs H_s^*-\frac12
(\varepsilon_0 \bs E_s\.\bs E_s^*+
 \mu_0          \bs H_s\.\bs H_s^*)\bs n\Bigr]\td\sigma \\[3mm]&
 =-\dfrac14\,\text{Re}\displaystyle\oint_{S_\infty}
\bigl[\varepsilon_0 \bs E_s\.\bs E_s^*+
 \mu_0          \bs H_s\.\bs H_s^*\bigr]\bs n\td\sigma  \\[3mm] &
 =-\dfrac{\varepsilon_0}{4k^2}\displaystyle\oint_{4\pi}
(| \bs a_{\bs n}|^2 + | \bs b_{\bs n}|^2)\bs n\td\Omega_n
=-\dfrac{\varepsilon_0}{2k^2}\displaystyle\oint_{4\pi}
|\bs a_{\bs n}|^2\bs n\td\Omega_n, 
\end{array}
\end{equation}
which is the second equation in Eqs.~(12) of the main text.

Now let us focus on the surface integral of the orbital momentum density given by the right hand side of
Eq.~(3a) in the main text,
\begin{equation}
-c\doint_{S_\infty} \av{\bs M^o_{\text{mix}}}\td\sigma -c\doint_{S_\infty} \av{\bs M^o_{\text{sca}}}  \td\sigma.  \label{Moabapp}
\end{equation}
For a monochromatic optical field,
the period-averaged momentum density $\av{\bs M}$ reads
\cite{bliokh2013,bliokh2014,beksh2015,berry2009,bliokh2015}
\begin{subequations}\label{PoPsapp}
\begin{eqnarray} \label{PoPs0}
\av{\bs M}  &\!=&\! \dfrac1{2c^2}\,\text{Re}\,[\bs E\x\bs H^*]=\av{\bs M^o}+\av{\bs M^s},
\end{eqnarray}
with
\begin{eqnarray}
\av{\bs M^o}&\!=&\! \dfrac{\varepsilon_0}{4\,\omega}\,\text{Im}\,[(\nabla\bs E)\.\bs E^* + Z_0^2(\nabla \bs H)\.\bs H^*], \mspace{30mu}\\
\av{\bs M^s}&\!=&\! \dfrac{1}2\,\nabla\x\av{\bs S},  \qquad
\av{\bs S}  = \dfrac{\varepsilon_0}{4\,\omega}\,\text{Im}\,[\bs E^*\x\bs E + Z_0^2\,\bs H^*\x\bs H],
\end{eqnarray}
\end{subequations}
where $\omega$ and $Z_0=\sqrt{\mu_0/\varepsilon_0}$ are the angular frequency
and the wave impedance, respectively, and we adopt
the notation 
$[(\nabla\bs X)\.\bs Y^*\,]_i=(\pdi{i}X_j)Y_j^*$,
with the subscript denoting the Cartesian component and summation over repeated subscripts assumed.
The decomposition of the total fields Eq.~(\ref{ehtapp}) suggests that the orbital momentum density 
be written in the form
\begin{equation} \label{Mmixsca}
\begin{array}{lll}
\av{\bs M^o_{\text{tot}}}=\av{\bs M^o_{\text{inc}}}+\av{\bs M^o_{\text{sca}}}+\av{\bs M^o_{\text{mix}}}, \\[2mm]
\av{\bs M^o_{\text{mix}}}
=-\dfrac{\varepsilon_0}{4\omega}\,\text{Im}\,[(\nabla\bs E_{\text{inc}}^*)\.\bs E_{\text{sca}}
                                                -(\nabla\bs E_{\text{sca}})\.\bs E_{\text{inc}}^*
    + Z_0^2(\nabla \bs H_{\text{inc}}^*)\.\bs H_{\text{sca}}
    -Z_0^2(\nabla \bs H_{\text{sca}})\.\bs H_{\text{inc}}^*], \\[3mm]
\av{\bs M^o_{\text{sca}}}
=-\dfrac{\varepsilon_0}{4\omega}\,\text{Im}\,[(\nabla\bs E_{\text{sca}}^*)\.\bs E_{\text{sca}}
    + Z_0^2(\nabla \bs H_{\text{sca}}^*)\.\bs H_{\text{sca}}],
\end{array}
\end{equation}
Integrating over surface $S_\infty$ term by term, and using Jones' lemma and Eqs.~(\ref{ehincapp}--\ref{ehsapp}), one has 
\begin{subequations}
\begin{eqnarray}
\text{Im}\displaystyle\oint_{S_\infty}
(\nabla\bs E^*_{\text{inc}})\.\bs E_{\text{sca}}\td\sigma 
&=&-\text{Re}\displaystyle\oint_{4\pi} \td\Omega_u \,\dfrac{e^{ikr}}{r}\displaystyle\oint_{S_\infty}\bs u(\bs e_{\bs u}^{\,*}\.\bs a_{\bs n}) e^{-ik(\bs u\.\bs r)}\td\sigma \nonumber \\
&=&\dfrac{2\pi}{k}\,\text{Im}\displaystyle\oint_{4\pi}  \bigl[(\bs e_{\bs u}^{\,*}\.\bs a_{\bs u})-(\bs e_{\bs u}^{\,*}\.\bs a_{-\bs u})e^{2ikR}\bigr]\bs u\,\td\Omega_u , \\
\text{Im}\displaystyle\oint_{S_\infty}
(\nabla\bs E_{\text{sca}})\.\bs E^*_{\text{inc}}\td\sigma 
&=&\text{Re}\displaystyle\oint_{4\pi} \td\Omega_u \,\dfrac{e^{ikr}}{r}\displaystyle\oint_{S_\infty}\bs n(\bs a_{\bs n}\.\bs e_{\bs u}^{\,*}) e^{-ik(\bs u\.\bs r)}\td\sigma \nonumber \\
&=&-\dfrac{2\pi}{k}\,\text{Im}\displaystyle\oint_{4\pi}  \bigl[(\bs e_{\bs u}^{\,*}\.\bs a_{\bs u})+(\bs e_{\bs u}^{\,*}\.\bs a_{-\bs u})e^{2ikR}\bigr]\bs u\,\td\Omega_u,
\end{eqnarray}\label{delEdotE}
\end{subequations}
followed by
\begin{equation}
\text{Im}\displaystyle\oint_{S_\infty}
\bigl[(\nabla\bs E^*_{\text{inc}})\.\bs E_{\text{sca}}
-(\nabla\bs E_{\text{sca}})\.\bs E_{\text{inc}}^*\bigr]\td\sigma =
\dfrac{4\pi}{k}\,\text{Im}\displaystyle\oint_{4\pi} (\bs e_{\bs u}^{\,*}\.\bs a_{\bs u})\bs u\,\td\Omega_u,
\end{equation}
and, similarly,
\begin{equation}
\text{Im}\displaystyle\oint_{S_\infty}Z_0^2
\bigl[(\nabla\bs H^*_{\text{inc}})\.\bs H_{\text{sca}}
-(\nabla\bs H_{\text{sca}})\.\bs H_{\text{inc}}^*\bigr]\td\sigma =
\dfrac{4\pi}{k}\,\text{Im}\displaystyle\oint_{4\pi} (\bs h_{\bs u}^{\,*}\.\bs b_{\bs u})\bs u\,\td\Omega_u,
\end{equation}
resulting in the first equation in Eq.~(14) of the main text,
\begin{equation}
\displaystyle\oint_{S_\infty}\av{\bs M^o_{\text{mix}}}\td\sigma
=-\dfrac{\pi\varepsilon_0}{k^2c}\,\text{Im}\displaystyle\oint_{4\pi}
\bigl[\,\bs e_{\bs u}^{\,*}\.\bs a_{\bs u}+\bs h_{\bs u}^{\,*}\.\bs b_{\bs u}\,\bigr]\bs u\,\td\Omega_u
=-\dfrac{2\pi\varepsilon_0}{k^2c}\,\text{Im}\displaystyle\oint_{4\pi}
\bigl(\bs e_{\bs u}^{\,*}\.\bs a_{\bs u}\bigr)\bs u\,\td\Omega_u. \label{intMmix}
\end{equation}
In deriving Eq.~(\ref{delEdotE}), we keep the leading term only so that
\begin{equation} \label{delEsca}
\nabla \bs E_{\text{sca}}=\nabla\left[\bs a_{\bs n} \,\dfrac{e^{ikr}}{kr}\right]=(ik\,\bs n\, \bs a_{\bs n}) \,\dfrac{e^{ikr}}{kr}, \qquad
\nabla \bs H_{\text{sca}}=\nabla\left[\bs b_{\bs n} \,\dfrac{e^{ikr}}{kr}\right]=(ik\,\bs n\, \bs b_{\bs n}) \,\dfrac{e^{ikr}}{kr},
\end{equation}
where $\nabla\bs a_{\bs n}$ and $\nabla\bs b_{\bs n}$ are neglected since they gives $1/r^2$ terms that do not contribute to the optical force.
Care should be taken in dealing with terms proportional to $1/r^2$ for the case of optical torque, which adds considerably to the complexity, as can be seen in Section S5.

For the second term in (\ref{Moabapp}), using Eqs. (\ref{Mmixsca}) and (\ref{delEsca}), one has
\begin{equation}\begin{array}{lll}
\displaystyle\oint_{S_\infty}\av{\bs M^o_{\text{sca}}}\td\sigma
&=-\dfrac{\varepsilon_0}{4\omega}\,\text{Im}\displaystyle\oint_{S_\infty}\bigl[(\nabla\bs E^*_{\text{sca}})\.\bs E_{\text{sca}}
+Z_0^2(\nabla\bs H^*_{\text{sca}})\.\bs H_{\text{sca}}\bigr]\td\sigma \\[2mm]
&=\dfrac{\varepsilon_0}{4k^2c}\,\text{Re}\displaystyle\oint_{4\pi}(|\bs a_{\bs n}|^2+|\bs b_{\bs n}|^2)\bs n\,\td\Omega_n 
=\dfrac{\varepsilon_0}{2k^2c}\,\text{Re}\displaystyle\oint_{4\pi}|\bs a_{\bs n}|^2\bs n\,\td\Omega_n.
\end{array}\label{intMsca}
\end{equation}
which is the second equation of Eq.~(14) in the main text.

Finally, \revlina{let us look at}
the surface integral of the orbital momentum current density given by the right hand side of
Eq.~(3b) in the main text, which, based on the conservation of canonical momentum in free-space
(see, discussion in Appendix \ref{appa}), is written as
\begin{equation}
-\doint_{S} \av{\bTTo}\.\bs n\td\sigma=-\doint_{S_\infty}\! \av{\bTTo}\.\bs n\td\sigma.   \label{TToab}
\end{equation}
\revlina{Decompose}
$$
\av{\bTTo}=\dfrac{1}{4\omega}\text{Im}\,
\bigl[(\nabla\bs E)\x\bs H^*+(\nabla\bs H^*)\x\bs E\bigr]
$$
into
$$
\av{\bTTo}= \av{\bTToi}+ \av{\bTTox}+\av{\bTTos},
$$
with
\begin{equation}\label{ToxTos}
\begin{array}{ll}
\revlina{\av{\bTToi}=\dfrac{1}{4\omega}\text{Im}\,
\bigl[(\nabla\bs E_{\text{inc}})\x\bs H^*_{\text{inc}}+(\nabla\bs H_{\text{inc}}^*)\x\bs E_{\text{inc}}\bigr],} \\[3mm] \av{\bTTox}=\dfrac{1}{4\omega}\text{Im}\,
\bigl[(\nabla\bs E_{\text{sca}})\x\bs H^*_{\text{inc}}-(\nabla\bs E_{\text{inc}}^*)\x\bs H_{\text{sca}}
+(\nabla\bs H^*_{\text{inc}})\x\bs E_{\text{sca}}-(\nabla\bs H_{\text{sca}})\x\bs E_{\text{inc}}^*\bigr]  \\[3mm]
\av{\bTTos}=\dfrac{1}{4\omega}\text{Im}\,
\bigl[(\nabla\bs E_{\text{sca}})\x\bs H^*_{\text{sca}}+(\nabla\bs H_{\text{sca}}^*)\x\bs E_{\text{sca}}\bigr]  .
\end{array}
\end{equation}
By Jones' lemma, the integral of $\av{\bTTox}\.\bs n$ over $S_\infty$ can be evaluated term by term, resulting in
\begin{subequations}
\begin{eqnarray}
\text{Im}\displaystyle\oint_{S_\infty}
\bigl[(\nabla\bs E_{\text{sca}})\x\bs H^*_{\text{inc}}\bigr]\.\bs n\td\sigma 
&=&\text{Im}\displaystyle\oint_{4\pi} \td\Omega_u \,\dfrac{ike^{ikR}}{Z_0kR}\displaystyle\oint_{S_\infty}
(\bs n\,\bs a_{\bs n}\x\bs h_{\bs u}^*)\.\bs n\, e^{-ik(\bs u\.\bs R)}\td\sigma \nonumber \\
&=&-\dfrac{2\pi}{Z_0k}\,\text{Im}\displaystyle\oint_{4\pi} \td\Omega_u
\bigl[(\bs u\,\bs a_{\bs u}\x\bs h_{\bs u}^*)\.\bs u
-(\bs u\,\bs a_{-\bs u}\x\bs h_{\bs u}^*)\.\bs u\,e^{2ikR}\bigr] \nonumber \\
&=&-\dfrac{2\pi}{Z_0k}\,\text{Im}\displaystyle\oint_{4\pi}
\bigl[( \bs a_{\bs u}\.\bs e_{\bs u}^*)
-(\bs a_{-\bs u}\.\bs e_{\bs u}^*)\,e^{2ikR}\bigr]\bs u\, \td\Omega_u,   \\
-\text{Im}\displaystyle\oint_{S_\infty}
\bigl[(\nabla\bs E_{\text{inc}}^*)\x\bs H_{\text{sca}}\bigr]\.\bs n\td\sigma
&=&-\dfrac{2\pi}{Z_0k}\,\text{Im}\displaystyle\oint_{4\pi}
\bigl[(\bs a_{\bs u}\.\bs e_{\bs u}^*)
-(\bs a_{-\bs u}\.\bs e_{\bs u}^*)\,e^{2ikR}\bigr]\bs u\,\td\Omega_u, \\
\text{Im}\displaystyle\oint_{S_\infty}
\bigl[(\nabla\bs H_{\text{inc}}^*)\x\bs E_{\text{sca}}\bigr]\.\bs n\td\sigma 
&=&-\dfrac{2\pi}{Z_0k}\,\text{Im}\displaystyle\oint_{4\pi}
\bigl[(\bs a_{\bs u}\.\bs e_{\bs u}^*)
+(\bs a_{-\bs u}\.\bs e_{\bs u}^*)\,e^{2ikR}\bigr] \bs u\,\td\Omega_u, \\
-\text{Im}\displaystyle\oint_{S_\infty}
\bigl[(\nabla\bs H_{\text{sca}})\x\bs E_{\text{inc}}^*\bigr]\.\bs n\td\sigma 
&=&-\dfrac{2\pi}{Z_0k}\,\text{Im}\displaystyle\oint_{4\pi}
\bigl[(\bs a_{\bs u}\.\bs e_{\bs u}^*)
+(\bs a_{-\bs u}\.\bs e_{\bs u}^*)\,e^{2ikR}\bigr] \bs u\,\td\Omega_u.
\end{eqnarray}
\end{subequations}
Putting together, one arrives at
\begin{equation}
\begin{array}{lll}
\doint_{S_\infty} \av{\bTTox}\.\bs n\td\sigma
=-\dfrac{2\pi\varepsilon_0}{k^2}\,\text{Im}\displaystyle\oint_{4\pi}
\bigl(\bs e_{\bs u}^{\,*}\.\bs a_{\bs u}\bigr)\bs u\,\td\Omega_u
=c\doint_{4\pi}\av{\bs M^o_{\text{mix}}}\td\sigma
\end{array}
\end{equation}
which is the first line of Eqs.~(16) in the main text.
Plugging in Eqs.~(\ref{ehsapp}) and (\ref{delEsca}) to the second line of (\ref{ToxTos}) leads to
\begin{equation}
\begin{array}{lll}
\doint_{S_\infty} \av{\bTTos}\.\bs n\td\sigma
=\dfrac{\varepsilon_0}{2k^2}\,\text{Re}\displaystyle\oint_{4\pi}|\bs a_{\bs n}|^2\bs n\,\td\Omega_n
=c\doint_{4\pi}\av{\bs M^o_{\text{sca}}}\td\sigma,
\end{array}
\end{equation}
which coincides with the second equation of Eqs.~(16) in the main text.

\vskip 20pt

\noindent
\section{The 
multipole moments versus
the partial wave expansion coefficients for the scattered fields
} 

Before proceeding to compute optical torque, we derive the relationship between the totally symmetric and traceless multipole moments
and the expansion coefficients for the scattered fields in terms of the vector spherical wave functions(VSWFs).

In the T-matrix method \cite{Tmatrix}, the scattered fields from any object can be written in the form
\begin{equation} \label{Esct}
\begin{array}{l}
\bs E_s=\dsum_{l=1}^{\infty}\dsum_{m=-l}^{l} i^{l+1}C_{ml}E_0
      \Bigl[\,a_{m,l}\bN_{ml}^{(3)}(k,\br)
            + b_{m,l}\bM_{ml}^{(3)}(k,\br)\,\Bigr]  \\[3mm]
\bs H_s=\dfrac{1}{Z_0} \dsum_{l=1}^{\infty}\dsum_{m=-l}^{l}  i^{l}C_{ml}E_0
      \Bigl[\,b_{m,l}\bN_{ml}^{(3)}(k,\br)
           + a_{m,l}\bM_{ml}^{(3)}(k,\br)\,\Bigr].
\end{array}
\end{equation}
where $Z_0=\sqrt{\mu_0/\varepsilon_0}$ is the background wave impedance, $E_0>0$ characterizes the incident field amplitude,
and
$$
C_{ml}=\left[\frac{(2l+1)}{l(l+1)} \frac{(l-m)!}{(l+m)!}\right]^{1/2}.
$$
The coefficients $a_{m,l}$ and $b_{m,l}$ are known as the expansion coefficients in terms of the VSWFs, or partial wave coefficients.
The VSWFs $\bM^{(3)}_{ml}$ and $\bN^{(3)}_{ml}$ describe the multipole fields \cite{stratton,rose}. They
are given by, see, {\it e.g.}, \cite{jack2005,stratton,bohren1998},
\begin{equation}
\begin{array}{ll}
\bM^{(3)}_{ml}(k,\br)&\!\!=
      \bigl[\,i\pi_{ml}(\cos\theta)\etheta-\tau_{ml}(\cos\theta)\ephi\,\bigr]
        \dfrac{\xi_l(kr)}{kr}e^{im\phi},  \\[3mm]
\bN^{(3)}_{ml}(k,\br) &\!\!=
      \bigl[\,\tau_{ml}(\cos\theta)\etheta+i\pi_{ml}(\cos\theta)\ephi\,\bigr]
          \dfrac{\xi'_l(kr)}{kr}e^{im\phi}  + \er l(l+1) P^m_l(\cos\theta) \dfrac{\xi_l(kr)}{(kr)^2}e^{im\phi},
\end{array}
\end{equation}
where the two auxiliary functions, $\pi_{ml}(\cos\theta)$ and
$\tau_{ml}(\cos\theta)$, are defined by
\begin{equation}
\pi_{ml}(\cos\theta)=\frac{m}{\sin\theta} P^m_l(\cos\theta), \quad
\tau_{ml}(\cos\theta)=\frac{\td}{\td\theta} P^m_l(\cos\theta),
\end{equation}
with $P^m_l(x)$ denoting the associated Legendre function of first kind and $\xi_l(x)$ the Riccati Hankel function \cite{bohren1998}.
The far fields can therefore be obtained using the asymptotic expansions of the Riccati Hankel function $\xi_l(x)$.

On the other hand, the far fields scattered from any object can be written in the form  of Eq. (\ref{ehsapp}). 
The amplitudes of the scattered fields are the sum of contribution from each individual electric and magnetic multipoles,
\begin{equation}\label{anbnapp}
\begin{array}{lll}
\bs a_{\bs n}=\dsum_{l=1}^{\infty}\left[\bs a_{\text{elec}}^{(l)}+\bs a_{\text{mag}}^{(l)}\right], & &
\alpha_{\bs n}=\dsum_{l=1}^{\infty}\alpha^{(l)}, \\[3mm]
\bs b_{\bs n}=\dsum_{l=1}^{\infty}\left[\bs b_{\text{elec}}^{(l)}+\bs b_{\text{mag}}^{(l)}\right], &\quad&
\beta_{\bs n}=\dsum_{l=1}^{\infty}\beta^{(l)}.
\end{array}
\end{equation}
where $\bs a_{\text{elec}}^{(l)}$ and $\bs a_{\text{mag}}^{(l)}$
[$\bs b_{\text{elec}}^{(l)}$ and $\bs b_{\text{mag}}^{(l)}$]
describe the transverse amplitudes of the scattered electric (magnetic) far-field
from an electric and a magnetic $2^l$-pole, respectively,
and $\alpha_{\bs n}$ ($\beta_{\bs n}$) represent the
longitudinal electric (magnetic) far-field due to an electric (a magnetic) $2^l$-pole.
Based on the theory of multipole fields \cite{rose,stratton} and the irreducible tensor method \cite{silver},
the amplitudes can be written as
\begin{equation}\label{alblapp}
\begin{array}{lllll}
\bs a_{\text{elec}}^{(l)}
&=\dfrac{i}{4\pi \varepsilon_0}\dfrac{(-ik)^{l+2}}{l\,!}\; \bs n\x \bigl(\bs n\x\bs{q}^{\,(l)}_{\text{elec}}\bigr), \mspace{20mu} &
&\bs a_{\text{mag}}^{(l)}
=\dfrac{i}{4\pi \varepsilon_0c}\dfrac{(-ik)^{l+2}}{l\,!}\;\bs n\x\bs{q}^{\,(l)}_{\text{mag}}, \\[3mm]
\bs b^{(l)}_{\text{elec}}
&=\bs n\x \bs a^{(l)}_{\text{elec}} , \mspace{20mu} & &
\bs b^{(l)}_{\text{mag}}
\,=\,\bs n\x \bs a^{(l)}_{\text{mag}}, \\[3mm]
\alpha^{(l)}
&=-\dfrac{(l+1)}{4\pi\varepsilon_0}\,\dfrac{(-ik)^{l+2}}{l\,!}\;\bigl(\bs n\.\bs q^{\,(l)}_{\text{elec}}\bigr) , \mspace{20mu} &
&\beta^{(l)}
=-\dfrac{(l+1)}{4\pi\varepsilon_0c}\,\dfrac{(-ik)^{l+2}}{l\,!}\;\bigl(\bs n\.\bs q^{\,(l)}_{\text{mag}}\bigr),
\end{array}
\end{equation}
with
\begin{equation}
\begin{array}{lll}
\bs q^{\,(l)}_{\text{elec\,(mag)}}
=\bnn{(l-1)}\;\tc{l-1}\bOO{\text{elec\,(mag)}}{(l)}, \mspace{30mu} \text{and} \quad
\bnn{(j)}=\overbrace{\bs{n}\,\bs{n}\cdots\bs{n}}^{j} 
\end{array}
\label{qvecapp}
\end{equation}
denoting the $j$-fold tensor product of $\bs n$, yielding a symmetric tensor of rank $j$.
The symbol $\tc{m}$ represents the multiple tensor contraction between two tensors of ranks $l$ and $l'$,
reslting in a tensor of rank $(l+l'-2m)$ defined by, 
\begin{equation}\label{AtcBapp}
\begin{array}{ll}
\stackrel{\text{\scriptsize$\boldsymbol{\leftrightarrow}$}}
         {\boldsymbol{\mathbb{A}}}^{\text{\raisebox{-1.8ex}{$(l)$}}}
         \tc{m}
\stackrel{\text{\scriptsize$\boldsymbol{\leftrightarrow}$}}
         {\boldsymbol{\mathbb{B}}}^{\text{\raisebox{-1.8ex}{$(l')$}}}
=
\mathbb{A}^{(l)}_{i_1\,i_2\,\cdots\,i_{l-m}\,\revlin{{k_1\,k_2\,\cdots\,k_{m-1}\,k_m}}}\,
\mathbb{B}^{(l')}_{\revlin{{k_m\,k_{m-1}\,\cdots\,k_2\,k_1}}\,j_{m+1}\,\cdots\,j_{l'-1}\,j_{l'}}, \quad 0\le m\le\min\,[\,l,l'\,],
\end{array}
\end{equation}
with the summation over repeated indices assumed and the subscript indexing the Cartesian component. 
The totally symmetric and traceless \cite{apple1989} rank-$l$ tensors
$\bOO{\text{elec}}{(l)}$ and $\bOO{\text{mag}}{(l)}$ represent the electric and magnetic $2^l$-pole moments induced on the particle.
The lower order cases with $l=1$, 2, 3, 4, and 5 correspond to the dipole,
quadrupole, octupole, hexadecapole, and dotriacontapole moments, respectively.
It is noted that the leading term $\alpha^{(l)}$ [\,$\beta^{(l)}$\,] of
the longitudinal components for electric (magnetic) far field comes
solely from the electric (magnetic) multipoles $\bOO{\text{elec}}{(l)}$ [\,$\bOO{\text{mag}}{(l)}$\,],
while the transverse amplitudes $\bs a_{\bs n}$ and $\bs b_{\bs n}$ depend on both electric and magnetic multipoles.

A comparison between $\bs E_{\text{sca}}$ and  $\bs E_s$, given by Eqs.~(\ref{ehsapp}) 
and (\ref{Esct}), respectively,
at large distance from the scatterer produces the relationship between
and the partial wave expansion coefficients $a_{m,l}$ and $b_{m,l}$ 
and the totally symmetric and traceless \cite{apple1989} electric and magnetic $2^l$-pole moments,
$\bOO{\text{elec}}{(l)}$ and $\bOO{\text{mag}}{(l)}$, whose
$3^l$ tensor elements can be expressed in terms of $2l+1$ independent components.

For lower order $l\le4$, the $2l+1$ independent components of the electric order
$l$ multipole ($2^l$-pole) moment $\bOO{\text{elec}}{(l)}$ are exemplified in terms of $a_{m,l}$ below.
\begin{subequations} \label{Ol}
\begin{equation}
\begin{array}{lll}
\mathbb O_{1}^{(1)}=p_1=c_1(a_{-1,1}-a_{1,1}), & \;\;
\mathbb O_{2}^{(1)}=p_2=-ic_1(a_{-1,1}+a_{1,1}),  & \;\;
\mathbb O_{3}^{(1)}=p_3= -\sqrt2c_1a_{0,1},
\end{array}
\end{equation}
for the 3 independent components of electric dipole moment  $\bOO{\text{elec}}{(1)}=\bs p$ with $l=1$,
\begin{equation}
\begin{array}{llll}
\mathbb O_{11}^{(2)}=Q_{11}=c_2(3 a_{-2,2} - \sqrt6 a_{0,2} + 3 a_{2,2}), &\quad &
\mathbb O_{12}^{(2)}=Q_{12}=-3ic_2(a_{-2,2} -a_{2,2}), \\[2mm]
\mathbb O_{13}^{(2)}=Q_{13}=-3c_2(a_{-1,2} -a_{1,2}),  & &
\mathbb O_{22}^{(2)}=Q_{22}=-c_2(3 a_{-2,2} + \sqrt6 a_{0,2} + 3 a_{2,2}),\\[2mm]
\mathbb O_{23}^{(2)}=Q_{23}=3ic_2(a_{-1,2} +a_{1,2}),
\end{array}
\end{equation}
for the 5 independent components of electric quadrupole moment $\bOO{\text{elec}}{(2)}=\bQ$ with $l=2$,
\begin{equation}
\begin{array}{lll}
\mathbb O_{111}^{(3)}=3c_3(\sqrt{15}a_{-3,3}-3a_{-1,3}+3a_{1,3}-\sqrt{15}a_{3,3}), &\quad&
\mathbb O_{112}^{(3)}=-3ic_3(\sqrt{15}a_{-3,3}-a_{-1,3}-a_{1,3}+\sqrt{15}a_{3,3}),\\[2mm]
\mathbb O_{113}^{(3)}=-3c_3(\sqrt{10}a_{-2,3}-2\sqrt3a_{0,3}+\sqrt{10}a_{2,3}), & &
\mathbb O_{122}^{(3)}=-3c_3(\sqrt{15}a_{-3,3}+a_{-1,3}-a_{1,3}-\sqrt{15}a_{3,3}),\\[2mm]
\mathbb O_{123}^{(3)}=3ic_3(\sqrt{10}a_{-2,3}-\sqrt{10}a_{2,3}), & &
\mathbb O_{222}^{(3)}=3ic_3(\sqrt{15}a_{-3,3}+3a_{-1,3}+3a_{1,3}+\sqrt{15}a_{3,3}),\\[2mm]
\mathbb O_{223}^{(3)}=3c_3(\sqrt{10}a_{-2,3}+2\sqrt3a_{0,3}+\sqrt{10}a_{2,3}),
\end{array}
\end{equation}
for the 7 independent components of electric octupole moment $\bOO{\text{elec}}{(3)}$ with $l=3$, and
\begin{equation}\mspace{-10mu}
\begin{array}{lll}
\mathbb O_{1111}^{(4)}=3c_4 (5 \sqrt{14} a_{-4,4} - 10 \sqrt2 a_{-2,4} + 6 \sqrt5 a_{0,4} - 10\sqrt2 a_{2,4}+ 5 \sqrt{14} a_{4,4}), \\[2mm]
\mathbb O_{1112}^{(4)}=-15\sqrt2\,ic_4 (\sqrt{7} a_{-4,4} -  a_{-2,4} +  a_{2,4}+ \sqrt{7} a_{4,4}), & \mspace{-125mu}
\mathbb O_{1113}^{(4)}=-15c_4 (\sqrt{7} a_{-3,4} - 3 a_{-1,4} + 3 a_{1,4} - \sqrt{7} a_{3,4}), \\[2mm]
\mathbb O_{1122}^{(4)}=-3c_4 (5\sqrt{14} a_{-4,4} - 2\sqrt5 a_{0,4} + 5\sqrt{14} a_{4,4}), &  \mspace{-125mu}
\mathbb O_{1123}^{(4)}=-3c_4(\sqrt{7} a_{-3,4} -  a_{-1,4} - a_{1,4} + \sqrt{7} a_{3,4}), \\[2mm]
\mathbb O_{1222}^{(4)}=15\sqrt2\,ic_4(\sqrt{7} a_{-4,4} + a_{-2,4} - a_{2,4} - \sqrt{7} a_{4,4}), &  \mspace{-125mu}
\mathbb O_{1223}^{(4)}=15c_4(\sqrt{7} a_{-3,4} + a_{-1,4} - a_{1,4} - \sqrt{7} a_{3,4}), \\[2mm]
\mathbb O_{2222}^{(4)}=3c_4 (5 \sqrt{14} a_{-4,4} + 10 \sqrt2 a_{-2,4} + 6 \sqrt5 a_{0,4} + 10\sqrt2 a_{2,4}+ 5 \sqrt{14} a_{4,4}), \\[2mm]
\mathbb O_{2223}^{(4)}=-15ic_4 (\sqrt{7} a_{-3,4} + 3 a_{-1,4} + 3 a_{1,4} + \sqrt{7} a_{3,4}),
\end{array}\mspace{-15mu}
\end{equation}
\end{subequations}
for the 9 independent components of electric hexadecapole moment $\bOO{\text{elec}}{(4)}$ with $l=4$,
where
$$
c_l=\dfrac{\sqrt{2l+1}}{l+1}\,\dfrac{(-i)^l 4\pi\varepsilon_0}{k^{l+2}}\,E_0.
$$
It is noted that the coefficients in front of $a_{\pm m,l}$ have the same amplitude, due to the introduction of $C_{ml}$ in Eq. (\ref{Esct}).

The magnetic multipole moments $\bOO{\text{mag}}{(l)}$ are obtained by replacing $a_{m,l}$ with $b_{m,l}$ in
Eq.~(\ref{Ol}) and multiplying the results by $-ic$, with $c$ being the speed of light.

\vskip 20pt

\noindent
\section{Optical torque} 

In this section, we derive the expressions for the extinction and recoil torques,
the surface integrals of optical angular momentum (AM) $\av{\bs J_{\text{mix}}}$ and $\av{\bs J_{\text{sca}}}$,
and the surface integrals of orbital and spin AM current density $\av{\bKKo}$ and $\av{\bKKs}$,
namely,  Eqs. (25), (26), (29) and (30) in the main text.

Decomposition of the total fields Eq.~(\ref{ehtapp}) suggests that 
$\langle\bKK\rangle=-\bs r\x\langle\bTT\rangle$
can be cast into,
\begin{equation}\label{tonrxT0}
\av{\bKK} = \av{\bKK_{\text{inc}}}+\av{\bKK_\text{mix}}+\av{\bKK_{\text{sca}}},
\end{equation}
where $\av{\bKK_{\text{inc}}}=-\bs r\x\av{\bTT_{\text{inc}}}$ involves the incident fields only,
$\av{\bKK_{\text{sca}}}=-\bs r\x\av{\bTT_{\text{sca}}}  $ depends solely on the scattered fields, while
$\av{\bKK_{\text{mix}}}=-\bs r\x\av{\bTT_{\text{mix}}} $ includes all the rest (mixed) terms, with
$\av{\bTT_{\eta}}$ given by Eqs.~(\ref{TiTmTs}) for  $\eta =\text{inc}$, $\text{mix}$, and $\text{sca}$.
Correspondingly, based on the Maxwell stress tensor formalism \cite{jackson,zangwill,schwinger}
the optical torque $\av{\bs T}$ reads (see, also, discussion in Appendix \ref{appa})
\begin{equation}\label{tonrxT}
\av{\bs T}= - \oint_{S_\infty}\av{\bKK}\. \bs n\td \sigma=
\av{\bs T_{\text{inc}}}+\av{\bs T_\text{mix}}+\av{\bs T_{\text{sca}}},
\end{equation}
where
\begin{equation}\label{tonrxT1}
\begin{array}{l}
\av {\bs T_{\text{inc}}}
 =-\displaystyle\oint_{S_\infty}\!\! \av{\bKK_{\text{inc}}}\.\bs n\td\sigma,\quad
 \av {\bs T_{\text{mix}}}
 =-\displaystyle\oint_{S_\infty}\!\! \av{\bKK_{\text{mix}}}\.\bs n\td\sigma, \quad
 \av {\bs T_{\text{sca}}}
 =-\displaystyle\oint_{S_\infty}\!\! \av{\bKK_{\text{sca}}}\.\bs n\td\sigma.
\end{array}
\end{equation}
The integration of $\av{\bKK_\text{inc}}$ gives no contribution to the time-averaged optical torque
since it is divergence-free,
viz,
\begin{equation}
\av {\bs T_{\text{inc}}}
=-\displaystyle\oint_{S_\infty}\!\! \av{\bKK_{\text{inc}}}\.\bs n\td\sigma
= \displaystyle\oint_{S_\infty}\bs r\x\bigl[\av{\bTT_{\text{inc}}}\.\bs n\bigr]\td\sigma =0.
\end{equation} 

The extinction torque $\av {\bs T_{\text{mix}}}$ consists of
the electric and magnetic parts, denoted by $\av{\bs T^{\text{\,e}}_{\text{mix}}}$ and $\av{\bs T^{\text{\,m}}_{\text{mix}}}$,
\begin{equation}
\av {\bs T_{\text{mix}}}
=-\displaystyle\oint_{S_\infty}\!\!\av{\bKK_{\text{mix}} }\.\bs n\td\sigma
=\displaystyle\oint_{S_\infty}\bs r\x\bigl[\av{\bTT_{\text{mix}}}\.\bs n\bigr]\td\sigma
=\av{\bs T^{\text{\,e}}_{\text{mix}}} + \av{\bs T^{\text{\,m}}_{\text{mix}}}.
\end{equation}
where
\begin{equation}
\begin{array}{ll}
\av{\bs T^{\text{\,e}}_{\text{mix}}}
=\dfrac{\varepsilon_0}2\,\text{Re}\displaystyle\oint_{S_\infty}
R\;\bs n\x\bigl[(\bs E_{\text{inc}}^{\,*}\bs E_{\text{sca}}+\bs E_{\text{sca}}\bs E_{\text{inc}}^{\,*})\. \bs n\bigr]\td\sigma,  \\[3mm]
\av{\bs T^{\text{\,m}}_{\text{mix}}}
=\dfrac{\mu_0}2\,\text{Re}\displaystyle\oint_{S_\infty}
R\;\bs n\x\bigl[(\bs H_{\text{inc}}^{\,*}\bs H_{\text{sca}}+\bs H_{\text{sca}}\bs H_{\text{inc}}^{\,*})\. \bs n\bigr]\td\sigma.
\end{array}
\end{equation}
It then follows from the expressions for the incident and scattered far-fields,
Eqs.~(\ref{ehincapp}) and (\ref{ehsapp}), that
\begin{equation} \begin{array}{rll}
\av{\bs T^{\text{\,e}}_{\text{mix}}}&=&\bs t_1+\bs t_2, \\[2mm]
\bs t_1
&=&
\dfrac{\varepsilon_0}{2k^2}\, \text{Re}\! \displaystyle\oint_{4\pi}\!\!\!\td\Omega_u\,
\dfrac{e^{ikR}}{R}\displaystyle\oint_{S_\infty}\!\!\!
(\bs n\x\bs e^{\,*}_{\bs u}\,)\alpha_{\bs n} \,e^{-ikR\,(\bs u\,\.\,\bs n)} \td\sigma_n, \\[3mm]
\bs t_2 &=& 
\dfrac{\varepsilon_0}{2k}\, \text{Re}\! \displaystyle\oint_{4\pi}\!\!\!\td\Omega_u\,
e^{ikR}\displaystyle\oint_{S_\infty}\!\!\!
(\bs n\x\bs a_{\bs n})(\bs e^{\,*}_{\bs u}\.\bs n)
  \,e^{-ikR\,(\bs u\,\.\,\bs n)} \td\sigma_n,
\end{array}
\end{equation}
which is Eq. (18) in the main text.
Here we use $\td\sigma_n$ to represent the area element for the integration over $S_\infty$.
It serves as a reminder of the fact that
$\bs n$ is the local outward unit normal of $\td \sigma_n$, while
the solid angle element $\td\Omega_u$ has a local outward unit normal denoted by $\bs u$.
The surface integral in the first term $\bs t_1$ of $\av{\bs T^{\text{\,e}}_{\text{mix}}}$ can be evaluated
using Jones Lemma Eq.~(\ref{jonesapp}) by setting $\bs G(\bs n)=\alpha_{\bs n}\,(\bs n\x\bs e^{\,*}_{\bs u})$,
resulting in
\begin{equation}
\bs t_1
= \dfrac{\pi\varepsilon_0}{k^3}\,\text{Re}\,\displaystyle\oint_{4\pi}
i\,(\alpha_{\bs u}+e^{2ikR}\,\alpha_{-\bs u}) \,
\bs h_{\bs u}^*\td\Omega_u. 
\end{equation}
The evaluation of the surface integral in the second term $\bs t_2$ of $\av{\bs T^{\text{\,e}}_{\text{mix}}}$ requires the extended Jones lemma (\ref{joneslinapp}),
which, as it involves the angular momentum operator $\boldsymbol{\hat{L}}$, requires the
explicit forms of the transverse and longitudinal far-field amplitudes
$\bs a_{\bs n}$,  $\bs b_{\bs n}$, $\alpha_{\bs n}$, and $\beta_{\bs n}$, given in Eqs. (\ref{anbnapp}-\ref{qvecapp}).

Substituting Eqs. (\ref{anbnapp}-\ref{qvecapp}) for $\bs a_{\bs n}$ and $\alpha_{\bs n}$ in $\av{\bs T^{\text{\,e}}_{\text{mix}}}$ and
using the extended Jones lemma Eq. (\ref{joneslinapp}), one arrives at, after lengthy algebra,
\begin{equation}
\begin{array}{llll}
\bs t_1&\!\!\!
=\dsum_{l=1}^{\infty} \bs t_1^{\;(l)}, \qquad
\bs t_2=\dsum_{l=1}^{\infty} \bs t_2^{\;(l)}, \qquad
 \\[4mm]
\bs t_1^{\;(l)}&\!\!\!
=-\dfrac{(l+1)k^{l-1}}{4\,l!}\,\text{Re}\displaystyle\oint_{4\pi}
           C^+_l\,\bigl[\,\bs u\.\bs q_{\text{elec}}^{\,(l)}\,\bigr]\bs h_{\bs u}^{\,*}\,\td\Omega_u, \\[4mm]
\bs t_2^{\;(l)}&\!\!\!
=\dfrac{k^{l-1}}{4\,l!}\,\text{Re}\displaystyle\oint_{4\pi}
           C^-_l\,\Bigl\{(l-1)\,\bs u\!\x\!\bigl[\bqq{(l)}{\text{elec}}\.\bs e_{\bs u}^{\,*}\bigr]
           +\bs e_{\bs u}^{\,*} \x  \bs q^{\,(l)}_{\text{elec}}   \Bigr\} \td\Omega_u
                \\[3mm]   & \mspace{3mu} {}
-\dfrac{k^{l-1}}{4\,c\,l!}\,\text{Re}\displaystyle\oint_{4\pi}
           C^+_l\,\Bigl\{
           (l+1)\bigl[\bs u\.\bs q^{\,(l)}_{\text{mag}}\bigr]\bs{e}_{\bs{u}}^{\,*}
           -(l-1)\,\bs u\!\x\!\bigl[\bqq{(l)}{\text{mag}}\.\bs h_{\bs u}^{\,*}\bigr]
           -\bs h_{\bs u}^{\,*} \x  \bs q^{\,(l)}_{\text{mag}} \Bigr\} \td\Omega_u ,
\end{array}
\end{equation}
where
\begin{equation}\label{Clpm}
C_l^{\pm}= (-i)^{l+1}\bigl[1\pm(-1)^l\,e^{2ikR}\bigr], \qquad
\left\{\begin{array}{ll}
\bqq{(l)}{\text{elec\,(mag)}}=\buu{(l-2)}\,\tc{l-2}\bOO{\text{elec\,(mag)}}{(l)}, & \quad \text{for $l>1$}, \\[2mm]
\bs q^{\,(l)}_{\text{elec\,(mag)}}=\buu{(l-1)}\,\tc{l-1}\bqq{(l)}{\text{elec\,(mag)}}.
\end{array}\right.
\end{equation}
So the electric contribution to the extinction torque reads
$$
\begin{array}{llll}
\av{\bs T^{\text{\,e}}_{\text{mix}}}\!\!
&=\dsum_{l=1}^{\infty} \av{\bs T^{\text{\,e}(l)}_{\text{mix}}}, \\[3mm]
\av{\bs T^{\text{\,e}(l)}_{\text{mix}}}\!\!
&=-\dfrac{(l+1)k^{l-1}}{4\,l!}\,\text{Re}\displaystyle\oint_{4\pi}
           C^+_l\,\bigl[\,\bs u\.\bs q_{\text{elec}}^{\,(l)}\,\bigr]\bs h_{\bs u}^{\,*}\,\td\Omega_u                 \\[3mm]   & \mspace{5mu} {}
+\dfrac{k^{l-1}}{4\,l!}\,\text{Re}\displaystyle\oint_{4\pi}
           C^-_l\,\Bigl\{(l-1)\,\bs u\!\x\!\bigl[\bqq{(l)}{\text{elec}}\.\bs e_{\bs u}^{\,*}\bigr]
           +\bs e_{\bs u}^{\,*} \x  \bs q^{\,(l)}_{\text{elec}}   \Bigr\} \td\Omega_u
                \\[3mm]   & \mspace{5mu} {}
-\dfrac{k^{l-1}}{4\,c\,l!}\,\text{Re}\displaystyle\oint_{4\pi}
           C^+_l\,\Bigl\{
           (l+1)\bigl[\bs u\.\bs q^{\,(l)}_{\text{mag}}\bigr]\bs{e}_{\bs{u}}^{\,*}
           -(l-1)\,\bs u\!\x\!\bigl[\bqq{(l)}{\text{mag}}\.\bs h_{\bs u}^{\,*}\bigr]
           -\bs h_{\bs u}^{\,*} \x  \bs q^{\,(l)}_{\text{mag}} \Bigr\} \td\Omega_u,
\end{array}
$$
the last term being the contribution to the scattered electric fields $\bs E_{\text{sca}}$ from the magnetic multipoles.
Similarly, the magnetic contribution to the extinction torque is
$$
\begin{array}{llll}
\av{\bs T^{\text{\,m}}_{\text{mix}}}\!\!
&=\dsum_{l=1}^{\infty} \av{\bs T^{\text{\,m}(l)}_{\text{mix}}}, \\[3mm]
\av{\bs T^{\text{\,m}(l)}_{\text{mix}}}\!\!
&=\dfrac{(l+1)k^{l-1}}{4\,c\,l!}\,\text{Re}\displaystyle\oint_{4\pi}
           C^+_l\,\bigl[\,\bs u\.\bs q_{\text{mag}}^{\,(l)}\,\bigr]\bs e_{\bs u}^{\,*}\,\td\Omega_u  \\[3mm]   & \mspace{3mu} {}
+\dfrac{k^{l-1}}{4\,c\,l!}\,\text{Re}\displaystyle\oint_{4\pi}
           C^-_l\,\Bigl\{(l-1)\,\bs u\!\x\!\bigl[\bqq{(l)}{\text{mag}}\.\bs h_{\bs u}^{\,*}\bigr]
           +\bs h_{\bs u}^{\,*} \x  \bs q^{\,(l)}_{\text{mag}}   \Bigr\} \td\Omega_u
                \\[3mm]   & \mspace{5mu} {}
+\dfrac{k^{l-1}}{4\,l!}\,\text{Re}\displaystyle\oint_{4\pi}
           C^+_l\,\Bigl\{
           (l+1)\bigl[\bs u\.\bs q^{\,(l)}_{\text{elec}}\bigr]\bs{h}_{\bs{u}}^{\,*}
           +(l-1)\,\bs u\!\x\!\bigl[\bqq{(l)}{\text{elec}}\.\bs e_{\bs u}^{\,*}\bigr]
           +\bs e_{\bs u}^{\,*} \x  \bs q^{\,(l)}_{\text{elec}} \Bigr\} \td\Omega_u.
\end{array}
$$
Summing up the electric and magnetic contriubutios for each $l$, one reaches
\begin{subequations} \label{Tmixint}
\begin{eqnarray}
\av{\bs T_{\text{mix}}^{(l)}}
&\!\!=&\!\! \av{\bs T^{\text{\,e}(l)}_{\text{mix}}}+\av{\bs T^{\text{\,m}(l)}_{\text{mix}}} , \\[1mm]
\av{\bs T_{\text{mix}}^{(l)}}
&\!\!=&\!\!
    \dfrac{k^{l-1}}{2\,l!}\,\text{Re} \doint_{4\pi}\!\!\td\Omega_u(-i)^{l+1}
    \Bigl[(l-1)\,\bs u\x\bigl(\bqq{(l)}{\text{elec}}\.\bs e_{\bs u}^{\,*}+\dfrac1c\,\bqq{(l)}{\text{mag}}\.\bs h_{\bs u}^{\,*}\bigr)
   +\bs{e}_{\bs{u}}^{\,*}\x \bs q_{\text{elec}}^{\,(l)}+\dfrac1c\,\bs{h}_{\bs{u}}^{\,*}\x \bs q_{\text{mag}}^{\,(l)}\Bigr].
\end{eqnarray}
\end{subequations}
Taking the integral over the solid angle gives
\begin{subequations}\label{Tmixapp}
\begin{eqnarray}
\av{\bs T_{\text{mix}}}
&\!\!=&\!\!
\dsum_{l=1}^{\infty}\av{\bs T_{\text{mix}}^{(l)}}, \\
\av{\bs T_{\text{mix}}^{(l)}}
&\!\!=&\!\!
\dfrac{1}{2\,l!}\,\text{Re}\bigl[(l-1)
 (\nabla^{(l-1)}\bs{E}_{\text{inc}}^*\,)\tc{l-1}\bOO{\text{elec}}{(l)}-
 \bOO{\text{elec}}{(l)}\tc{l-1}(\nabla^{(l-1)}\bs{E}_{\text{inc}}^*)\bigr]\tc{2}\beps\nonumber\\
&&\!\! {} +
 \dfrac{1}{2\,l!}\,\text{Re}\bigl[(l-1)
 (\nabla^{(l-1)}\bs{B}_{\text{inc}}^*\,)\tc{l-1}\bOO{\text{mag}}{(l)}
 -\bOO{\text{mag}}{(l)}\tc{l-1}(\nabla^{(l-1)}\bs{B}_{\text{inc}}^*)\bigr]\tc{2}\beps,   \label{Tmixl}
\end{eqnarray}
\end{subequations}
where $\beps$ is the Levi-Civita antisymmetric tensor of rank 3 (see, e.g., \cite{zangwill}),
$\nabla^{(j)}\bs{V}$ denotes the $j$-fold gradient of a vector $\bs V$, yielding a rank-$(j+1)$ tensor given,
in Cartesian components, by
$$
\nabla^{(j)}\bs{V}=\partial_{i_1}\partial_{i_2}\cdots\partial_{i_j}V_i,
$$
and the multiple tensor contraction is defined in Eq. (\ref{AtcBapp}).
When going from Eqs. (\ref{Tmixint}) to (\ref{Tmixapp}), one has made use of
\begin{equation}\begin{array}{rll}
\doint_{4\pi}\!\!\td\Omega_u\,
\bs{e}_{\bs{u}}^{\,*}\x \bs q_{\text{elec}}^{\,(l)}
&=&\dfrac{i^{l-1}}{k^{l-1}}\,\Bigl[\,\bOO{\text{elec}}{(l)}\tc{l-1}
\bigl(\nabla^{(l-1)} \bs E_{\text{inc}}^*\bigr)
\Bigr]\tc{2}\beps, \\[3mm]
-\doint_{4\pi}\!\!\td\Omega_u\,\bs u\x\bigl[\bqq{(l)}{\text{elec}}\.\bs{e}_{\bs{u}}^{\,*}\bigr]
&=&\dfrac{i^{l-1}}{k^{l-1}}\,\Bigl[
\bigl(\nabla^{(l-1)} \bs E_{\text{inc}}^*\bigr)
\tc{l-1}\,\bOO{\text{elec}}{(l)}\Bigr]\tc{2}\beps,
\end{array}
\end{equation}
which can be derived by mathematical induction.

The recoil torque $\av {\bs T_{\text{sca}}}$ reads, with the use of Eq. (\ref{ehsapp}),
\begin{equation}\label{Trecoil}
\av {\bs T_{\text{sca}}}
= \dfrac12\,\text{Re}\displaystyle\oint_{S_\infty}\!\!
R\,\bigl[
\varepsilon_0\bs n\x(\bs E_{\text{sca}}\bs E_{\text{sca}}^{\,*}\.\bs n) +
\mu_0\bs n\x(\bs H_{\text{sca}}\bs H_{\text{sca}}^{\,*}\.\bs n) \bigr]\td\sigma_n
=\dfrac{\varepsilon_0}{2k^3}\,\text{Re}\displaystyle\oint_{4\pi}
\bigl(\alpha_{\bs n}^*\,\bs b_{\bs n}-\beta_{\bs n}^*\,\bs a_{\bs n}\bigr)\td\Omega_n.
\end{equation}
With the explicit forms of field amplitudes Eqs.~(\ref{anbnapp}-\ref{qvecapp}), it can be worked out that
\begin{subequations} \label{Tscaapp}
\begin{eqnarray}
\av{\bs T_{\text{sca}}}&=&\dsum_{l=1}^{\infty}\av{\bs T^{(l)}_{\text{sca}}}, \\
\av{\bs T^{(l)}_{\text{sca}}} &=&-
\dfrac{k^{2l+1}}{8\pi\varepsilon_0} \,\dfrac{2^l(l+1)}{(2l+1)!}\;
\text{Im}\bigl[\,
\bOO{\text{elec}}{(l)}\tc{l-1}\bOO{\text{elec}}{(l)*} 
+ \dfrac{1}{c^2}\,\bOO{\text{mag}}{(l)}\tc{l-1}\bOO{\text{mag}}{(l)*}\,\bigr]\tc{2}\beps,  \label{Tscal}
\end{eqnarray}
\end{subequations}
where use has been made of the following mathematical identities,
\begin{subequations}
\begin{eqnarray}
 \displaystyle\oint_{4\pi}  \bigl[\alpha^{(l)\,*}\,\bs b^{(l')}_{\text{elec}}\bigr]\td\Omega_n
&=&
\dfrac{i\,k^{2l+4}}{4\pi\varepsilon_0^2} \,\dfrac{2^l(l+1)}{(2l+1)!}\,\,
\bigl[\,\bOO{\text{elec}}{(l)} \tc{l-1} \bOO{\text{elec}}{(l)\,*} \,\bigr]\tc{2}\beps\;\delta_{l,l'}, \\[2mm]
- 
\displaystyle\oint_{4\pi}  \bigl[\beta^{(l)\,*}\,\bs a^{(l')}_{\text{mag}}\bigr]\td\Omega_n
&=&\dfrac{i\,k^{2l+4}}{4\pi\varepsilon_0^2\,c^2} \,\dfrac{2^l(l+1)}{(2l+1)!}\,\,
\bigl[\,\bOO{\text{mag}}{(l)} \tc{l-1} \bOO{\text{mag}}{(l)\,*} \,\bigr]\tc{2}\beps\;\delta_{l,l'}, \\[2mm]
\displaystyle\oint_{4\pi}  \bigl[\alpha^{(l)\,*}\,\bs b^{(l')}_{\text{mag}}\bigr]\td\Omega_n
&=&\displaystyle\oint_{4\pi} \bigl[\beta^{(l')\,*}\,\bs a^{(l)}_{\text{elec}}\bigr]^{*}\td\Omega_n \\[2mm]
&=& 
  \dfrac{k^{2l'+3}}{4\pi\varepsilon_0^2\,c} \,\dfrac{2^{l'} l'\,(l'+1)}{(2l'+1)!}\,\,
\bigl[\,\bOO{\text{elec}}{(l)} \tc{l-1} \bOO{\text{mag}}{(l+1)\,*} \,\bigr]\tc{2}\beps\;\delta_{l+1,l'} \nonumber \\[2mm] & & {}
+  \dfrac{k^{2l+3}}{4\pi\varepsilon_0^2\,c} \,\dfrac{2^l l\,(l+1)}{(2l+1)!}\,\,
\bigl[\,\bOO{\text{elec}}{(l)} \tc{l-1} \bOO{\text{mag}}{(l-1)\,*} \,\bigr]\tc{2}\beps\;\delta_{l-1,l'},
\end{eqnarray}
\end{subequations}
which follow Eqs. (\ref{anbnapp}-\ref{qvecapp}) and mathematical induction.

Equations (\ref{Tmixapp}) and (\ref{Tscaapp}) constitute the multipole expansion of optical torque.
It is noted that the recoil torque depend solely on the contraction of electric or magnetic multipoles of the same order $l$, contrary to
the recoil optical force, which comes from the coupling between the same kind of multipoles of adjacent orders,
and the coupling between the different kinds of multipoles of the same order \cite{chen2011}.
So in the case only a single multipole is excited on the particle,
the particle experiences both extinction and recoil torques, but it is not subject to recoil force, resulting in
a positive-definite optical force.

As a simplest example, consider the case with $l=1$, viz, a small particle with electric and magnetic dipole excitations.
The optical torque Eq.~(\ref{Tmixapp}) and (\ref{Tscaapp}) reduces to
\begin{equation}
\av{\bs T_{\text{dip}}}=
-\dfrac{1}{2}\,\text{Re}\bigl[\bOO{\text{elec}}{(1)}\bs{E}_{\text{inc}}^*+\bOO{\text{mag}}{(1)}\bs{B}_{\text{inc}}^*\bigr]\tc{2}\beps
-\dfrac{k^3}{12\pi\varepsilon_0}\,\text{Im}\bigl[\,
\bOO{\text{elec}}{(1)} \bOO{\text{elec}}{(1)*} 
+ \dfrac{1}{c^2}\,\bOO{\text{mag}}{(1)} \bOO{\text{mag}}{(1)*}\,\bigr]\tc{2}\beps,
\end{equation}
where $\bOO{\text{elec}}{(1)}$ and $\bOO{\text{mag}}{(1)}$ are tensors of rank-1 (vectors) and
represent the electric and magnetic dipole moments, usually denoted by $\bs p$ and $\bs m$ (see, {\it e.g.}, \cite{zangwill}), respectively.
So it follows straightforward that
\begin{equation}\label{Tdipapp}
\begin{array}{lll}
\av{\bs T_\text{dip}}
&\!\!=&\!\!\dfrac12\,\text{Re}(\bs p\x\bs E_{\text{inc}}^*)
   +\dfrac12\,\text{Re}(\bs m\x\bs B_{\text{inc}}^*)
   +\dfrac{k^3}{12\pi\varepsilon_0}\,\text{Im}(\bs p\x\bs p^{\,*})
   +\dfrac{\mu_0k^3}{12\pi}\,\text{Im}(\bs m\x\bs m^*),
\end{array}
\end{equation}
where use has been made of
\begin{equation}\label{vxw}
(\bs v\,\bs w)\tc{2}\beps=v_i\,w_j \,\epsilon_{jik}=\bs w\x\bs v =-\bs v\x\bs w,
\end{equation}
for vectors $\bs v$ and $\bs w$, by the definition of multiple tensor contraction Eq.~(\ref{AtcBapp}).
The last two terms in $\av{\bs T_\text{dip}}$ describe the recoil torque,
which cancels out the extinction torque (the first two terms).
This complete cancelation between the extinction torque and recoil torque occurs
on any non-absorbing spherical particle in arbitrary monochromatic optical fields \cite{jiang2015,nieto2015a,nieto2015b}.
If one keeps up to electric quadrupole, the additional term
$\av{\bs T_\text{e-quad}}$ follows Eqs.~(\ref{Tmixapp}) and (\ref{Tscaapp}), reading
\begin{equation}
\av{\bs T_\text{e-quad}}
=\dfrac1{4}\,\text{Re}\bigl[(\nabla \bs E^*_{\text{inc}})\.\bQ-\bQ\.(\nabla \bs E^*_{\text{inc}})\bigr]\tc{2}\beps
 - \dfrac{k^5}{80\pi\varepsilon_0}\,\text{Im}
               \bigl[ \bQ\.\bQ^{\text{\raisebox{-1.8ex}{$*$}}} \bigr]\tc{2}\beps,
\end{equation}
where the electric quadrupole moment $\bQ=\bOO{\text{elec}}{(2)}$ is a symmetric and traceless rank-2 tensor.
Written in a more familiar form, it becomes
\begin{equation}
\begin{array}{lll}
\av{\bs T_\text{e-quad}}
&=\dfrac1{2}\,\text{Re}\bigl[(\bQ\.\nabla)\x \bs E^*_{\text{inc}}
  +\dfrac1{2}\, \bQ\.(\nabla\x\bs E_{\text{inc}}^*) \bigr]
  + \dfrac{k^5}{80\pi\varepsilon_0}\,\text{Im}
               \bigl[\,\bs Q_x\x\bs Q_x^*+\bs Q_y\x\bs Q_y^*+\bs Q_z\x\bs Q_z^*\,\bigr] \\[3mm]
&= \dfrac1{2}\,\text{Re}\bigl[(\bQ\.\nabla)\x \bs E^*_{\text{inc}}\bigr]
  +\dfrac{\omega}{4}\,\text{Im}\bigl[\bQ\.\bs B_{\text{inc}}^*\bigr]
  + \dfrac{k^5}{80\pi\varepsilon_0}\,\text{Im}
               \bigl[\,\bs Q_x\x\bs Q_x^*+\bs Q_y\x\bs Q_y^*+\bs Q_z\x\bs Q_z^*\,\bigr]
\end{array}
\end{equation}
or, in Cartesian component form,
\begin{equation}
\begin{array}{lll}
\av{\bs T_\text{e-quad}}_j&= \dfrac1{2}\,\text{Re}\bigl[Q_{i_1i_2}\partial_{i_2} E^*_{i_3}\,\epsilon_{i_1i_3j}\bigr]
  \revlin{+}\dfrac{\omega}{4}\,\text{Im}\bigl[Q_{ji} B^*_{i}\bigr]
  + \dfrac{k^5}{80\pi\varepsilon_0}\,\text{Im}
               \bigl[Q_{i_1i_2}Q_{i_2i_3}^*\epsilon_{i_1i_3j}\bigr],
\end{array}
\end{equation}
where $E^*_{i}$ ($B^*_{i}$) denotes the $i$-th Cartesian component of incident electric (magnetic) field $\bs E^*_{\text{inc}}$ ($\bs B^*_{\text{inc}}$), whereas
$\bs Q_x=\be_x\.\bQ$, \ $\bs Q_y=\be_y\.\bQ$, \ $\bs Q_z=\be_z\.\bQ$, \ and use has been made of
\begin{equation}
\bQ\.(\nabla\x\bs E_{\text{inc}}^*)=-\bigl[(\bQ\.\nabla) \x\bs E_{\text{inc}}^*+\nabla\x(\bs E_{\text{inc}}^*\.\bQ)\bigr]
=\bigl[\bQ\.(\nabla \bs E^*_{\text{inc}})+(\nabla \bs E^*_{\text{inc}})\.\bQ\bigr]\tc{2}\beps,
\end{equation}
which follows
\begin{equation}
\begin{array}{lll}
\bigl[\bQ\.(\nabla \bs E^*_{\text{inc}})\bigr]\tc{2}\beps=-(\bQ\.\nabla) \x\bs E_{\text{inc}}^*, \qquad 
\bigl[(\nabla \bs E^*_{\text{inc}})\.\bQ\bigr]\tc{2}\beps=-\nabla\x(\bs E_{\text{inc}}^*\.\bQ), \\[3mm]
\bQ\.(\bs v\x\bs w)=\revlin{-}\bigl[(\bQ\.\bs v)\x\bs w+\bs v\x(\bs w\.\bQ)\bigr]
\end{array}
\label{wvQ}
\end{equation}
with the last line valid for a symmetric and traceless rank-2 tensor $\bQ$.

Now let us focus on the surface integral of total angular momentum, viz the right hand side in Eq. (3c) and, also, Eq. (28), in the main text
\begin{equation}
-c\doint_{S_\infty} \av{\bs J_{\text{mix}}}\td\sigma -c\doint_{S_\infty} \av{\bs J_{\text{sca}}}  \td\sigma,  \label{Jabapp}
\end{equation}
where the total AM densities are
\begin{equation}\label{Jmixsca}
\begin{array}{ll}
\av{\bs J_{\text{mix}}}= \av{\bs L_{\text{mix}}} + \av{\bs S_{\text{mix}}}
=\bs r\x \av{\bs M^o_{\text{mix}}} + \av{\bs S_{\text{mix}}}, \qquad 
\av{\bs J_{\text{sca}}}=\av{\bs L_{\text{sca}}} + \av{\bs S_{\text{sca}}}
=\bs r\x \av{\bs M^o_{\text{sca}}} + \av{\bs S_{\text{sca}}},
\end{array}
\end{equation}
with the orbital LM density $\av{\bs M^o}$ and spin AM density $\av{\bs S}$ reading
\begin{equation}\label{MoS}
\begin{array}{ll}
\av{\bs M^o_{\text{mix}}}=\dfrac{\varepsilon_0}{4\omega}\,\text{Im}\,[(\nabla\bs E_{\text{sca}})\.\bs E^*_{\text{inc}}
             - (\nabla\bs E^*_{\text{inc}})\.\bs E_{\text{sca}}
             + Z_0^2\,(\nabla \bs H_{\text{sca}})\.\bs H^*_{\text{inc}}
             - Z_0^2\,(\nabla \bs H^*_{\text{inc}})\.\bs H_{\text{sca}}],  \\[3mm]
\av{\bs M^o_{\text{sca}}}=\dfrac{\varepsilon_0}{4\omega}\,\text{Im}\,[(\nabla\bs E_{\text{sca}})\.\bs E^*_{\text{sca}}
             + Z_0^2\,(\nabla \bs H_{\text{sca}})\.\bs H^*_{\text{sca}}], \\[3mm]
\av{\bs S_{\text{mix}}}=\dfrac{\varepsilon_0}{2\omega}\,\text{Im}\,[\bs E^*_{\text{inc}}\x\bs E_{\text{sca}}
                    + Z_0^2\bs H^*_{\text{inc}}\x\bs H_{\text{sca}} ],\\[3mm]
\av{\bs S_{\text{sca}}}=\dfrac{\varepsilon_0}{4\omega}\,\text{Im}\,[\bs E^*_{\text{sca}}\x\bs E_{\text{sca}}
                    + Z_0^2\bs H^*_{\text{sca}}\x\bs H_{\text{sca}} ].
\end{array}
\end{equation}

By keeping only the leading terms, the gradient of fields follows from Eqs. (\ref{ehincapp}-\ref{anbneuhu})
\begin{equation}\label{delEHall}
\begin{array}{lll}
\nabla\bs E_{\text{sca}}=\Bigl(ik\,\bs n\,\bs a_{\bs n}+\dfrac{ik\,\alpha_{\bs n}\,\bs n\,\bs n}{kr}+\nabla\bs a_{\bs n}\Bigr) \dfrac{e^{ikr}}{kr}, \qquad
\nabla\bs E_{\text{inc}}=ik\doint_{4\pi}\bs u\,\be_{\bs u} \,e^{i k \bs u\.\bs r}\,\td\Omega_u,  \\[4mm]
\nabla\bs H_{\text{sca}}=\dfrac1{Z_0}\Bigl(ik\,\bs n\,\bs b_{\bs n}+\dfrac{ik\,\beta_{\bs n}\,\bs n\,\bs n}{kr}+\nabla\bs b_{\bs n}\Bigr) \dfrac{e^{ikr}}{kr}, \qquad
\nabla\bs H_{\text{inc}}=\dfrac{ik}{Z_0}\doint_{4\pi}\bs u\bh_{\bs u} \,e^{i k \bs u\.\bs r}\,\td\Omega_u,
\end{array}
\end{equation}
which keep higher order terms in comparison with (\ref{delEsca}), due to the presence of the moment arm
in orbital AM density $\av{\bs L}$.
Inserting (\ref{delEHall}) into Eqs. (\ref{Jmixsca}) and (\ref{MoS}) leads to
\begin{subequations}\label{Jmixsca1}
\begin{equation}
\left\{\begin{array}{ll}
-c\doint_{S_\infty}\av{\bs L_{\text{mix}}}\td\sigma = \bs j_1+\bs j_2 \\[3mm]
-c\doint_{S_\infty}\av{\bs S_{\text{mix}}}\td\sigma =\bs j_3, 
\end{array}\right. \qquad \text{and} \qquad
\left\{
\begin{array}{ll}
-c\doint_{S_\infty}\av{\bs L_{\text{sca}}}\td\sigma = \bs j_4 \\[3mm]
-c\doint_{S_\infty}\av{\bs S_{\text{sca}}}\td\sigma = \bs j_5
\end{array}\right.
\end{equation}
with
\begin{equation}\label{j1j2j3j4j5}
\begin{array}{lll}
\bs j_1 &=
-\dfrac{\varepsilon_0}{4k}\,\,\text{Re}\doint_{4\pi}\!\!\td\Omega_u\,
e^{ikR}\doint_{S_\infty}\!
(\bs n\x\bs u)\bigl(\bs a_{\bs n}\.\bs e_{\bs u}^{\,*}+\bs b_{\bs n}\.\bs h_{\bs u}^*\bigr)e^{-ikR\,(\bs u\,\.\,\bs n)}\td\sigma_n
    \\[3mm]
\bs j_2 &=- \dfrac{\varepsilon_0}{4k^2}\,\,\text{Im}\doint_{4\pi}\!\!\td\Omega_u\,
 e^{ikR}\doint_{S_\infty}\!\!
  \bs n\!\x\!\bigl[(\nabla\bs a_{\bs n})\.\bs e_{\bs u}^{\,*} + (\nabla\bs b_{\bs n})\.\bs h_{\bs u}^* \,\bigr]\,e^{-ikR\,(\bs u\,\.\,\bs n)}\td\sigma_n
      \\[3mm]
\bs j_3  &= -\dfrac{\varepsilon_0}{2 k}\,\text{Im}\doint_{4\pi}\td\Omega_u\,\dfrac{e^{ikR}}{kR}\doint_{S_\infty}
\bigl(\bs e_{\bs u}^{\,*}\x\bs a_{\bs n}+\bs h_{\bs u}^*\x\bs b_{\bs n}\bigr)\,e^{-ikR\,(\bs u\,\.\,\bs n)}\td\sigma_n,   \\[3mm]
\bs j_4  &=- \dfrac{\varepsilon_0}{2k^3}\,\,\text{Im}\doint_{4\pi}
\bs r\x\bigl[(\nabla\bs a_{\bs n})\.\bs a_{\bs n}^*\,\bigr]\,\td\Omega_n  \\[3mm]
\bs j_5 &= \dfrac{\varepsilon_0}{2k^3}\,\,\text{Im}\doint_{4\pi}
\bigl(\bs a_{\bs n}\x\bs a_{\bs n}^*\bigr)\td\Omega_n,
\end{array}
\end{equation}
\end{subequations}
where one keeps only the leading terms that make non-vanishing contribution to the integral.
Compared with Eq.~(\ref{delEsca}), the extra higher order terms in $\nabla\bs E_{\text{sca}}$ and  $\nabla\bs H_{\text{sca}}$ appearing in Eq.~(\ref{delEHall}) do not contribute to optical force,
but they contribute to optical torque through $\bs j_2$ and $\bs j_4$ in (\ref{j1j2j3j4j5}).
In deriving Eqs. (\ref{Jmixsca1}),
use has been made of 
$$\begin{array}{lll}
\nabla\bs n=\dfrac1r\,\bigl[\bI-\bs n\bs n\bigr], \qquad
(\nabla \bs b_{\bs n})\.\bs b_{\bs n}^*
=(\nabla\bs a_{\bs n})\.\bs a_{\bs n}^*, \qquad
\bs b_{\bs n}^*\x\bs b_{\bs n}
=\bs a_{\bs n}^*\x\bs a_{\bs n}.
\end{array}
$$

Using Jones lemma Eq.~(\ref{jonesapp}), the integral over $S_\infty$ in $\bs j_3$ can be taken to give
\begin{eqnarray}
\bs j_3  
&=&-\dfrac{\varepsilon_0\pi}{k^3}\,\text{Re}\doint_{4\pi}\bigl(\bs e_{\bs u}^{\,*}\x\bs a_{\bs u}-\bs e_{\bs u}^{\,*}\x\bs a_{-\bs u}e^{2ikR}
                                                +\bs h_{\bs u}^*\x\bs b_{\bs u}-\bs h_{\bs u}^*\x\bs b_{-\bs u}e^{2ikR}\bigr)\td\Omega_u \nonumber \\[1mm]
&=&-\dfrac{2\varepsilon_0\pi}{k^3}\,\text{Re}\doint_{4\pi} (\bs e_{\bs u}^{\,*}\x\bs a_{\bs u})\td\Omega_u
=-\dsum_{l=1}^{\infty}
 \dfrac{k^{l-1}}{2\,l!}\,\text{Re}\doint_{4\pi}\td\Omega_u
(-i)^{l+1}\Bigl[
\dfrac1c\, \bs e_{\bs u}^{\,*}\.\bs q_{\text{mag}}^{\,(l)}
- \bs h_{\bs u}^{\,*}\.\bs q_{\text{elec}}^{\,(l)}\Bigr]\bs u \label{j3}
\end{eqnarray}
where $\bs q_{\text{elec(mag)}}^{\,(l)}$ are given by (\ref{qvecapp}), and one has used
$$
\bs e_{\bs u}^{\,*}\x\bs a_{\pm\bs u}
=\pm\,\bs h_{\bs u}^*\x\bs b_{\pm\bs u}, \qquad
\bs e_{\bs u}^{\,*}\x\bs a_{\bs u}=\dsum_{l=1}^{\infty}
\dfrac{(-i)^{l+1}k^{l+2}}{4\pi\varepsilon_0\,l!}
\Bigl[\dfrac1c\,\bs e_{\bs u}^{\,*}\.\bs q_{\text{mag}}^{\,(l)}-\bs h_{\bs u}^{\,*}\.\bs q_{\text{elec}}^{\,(l)}\Bigr]\bs u.
$$

To compute the integrals over surface $S_\infty$ in $\bs j_1$ and $\bs j_2$, one has to resort to the extended Jones lemma Eq.~(\ref{joneslinapp}).
With the explicit expressions for $\bs a_{\bs n}$ and $\bs b_{\bs n}$, given in Eqs.~(\ref{anbnapp}-\ref{qvecapp})
in terms of multipole fields and after lengthy algebra, the terms associated with the electric multipole
($2^l$-pole) $\bOO{\text{elec}}{(l)}$ in $\bs j_1+\bs j_2$ can be worked out by mathematical induction, resulting in
\begin{equation}
\begin{array}{lll}
e^{ikR}\doint_{S_\infty}\!
(\bs n\x\bs u)\bigl(\bs a_{\text{elec}}^{(l)}\.\bs e_{\bs u}^{\,*} )\,e^{-ikR\,(\bs u\,\.\,\bs n)}\td\sigma_n
=
\dfrac{e^{ikR}}{ik}\doint_{S_\infty}\!\!
  \bs n\!\x\!\bigl[(\nabla\bs a_{\text{elec}}^{(l)})\.\bs e_{\bs u}^{\,*} \,\bigr]\,e^{-ikR\,(\bs u\,\.\,\bs n)}\td\sigma_n
 =  2\,C_l^+\bs A_{\text{elec}}^{(l)} ,   
\\[3mm]
e^{ikR}\doint_{S_\infty}\!
(\bs n\x\bs u)\bigl(\bs b_{\text{elec}}^{(l)}\.\bs h_{\bs u}^{\,*} )\,e^{-ikR\,(\bs u\,\.\,\bs n)}\td\sigma_n
=
\dfrac{e^{ikR}}{ik}\doint_{S_\infty}\!\!
  \bs n\!\x\!\bigl[(\nabla\bs b_{\text{elec}}^{(l)})\.\bs h_{\bs u}^{\,*} \,\bigr]\,e^{-ikR\,(\bs u\,\.\,\bs n)}\td\sigma_n
=2\,C_l^-\bs A_{\text{elec}}^{(l)},
\end{array}
\end{equation}
where $C_l^{\pm}$ 
are given by Eq.~(\ref{Clpm}), and,
\begin{equation} \label{Loelec}
\bs A_{\text{elec}}^{(l)}=-\dfrac{k^l}{4\,l!\,\varepsilon_0} 
  \Bigl\{ (l-1)\bs{u}\x\!\!\bigl[\bqq{(l)}{\text{elec}}\.\bs{e}_{\bs{u}}^{\,*}\bigr]
  -\bigl[\bs u\.\bs q_{\text{elec}}^{\,(l)}\bigr]\bs{h}_{\bs{u}}^{\,*} \Bigr\}.
\end{equation}
It follows that
\begin{equation}
\bs j_1+\bs j_2
=-\dfrac{2\varepsilon_0}{k}\,\,\text{Re}\dsum_{l=1}^{\infty}\doint_{4\pi}\td\Omega_u (-i)^{l+1} \bigl(\bs A_{\text{elec}}^{(l)}+\bs A_{\text{mag}}^{(l)}\,\bigr),
\end{equation}
where, according to the electric-magnetic duality \cite{zangwill}, the terms involving the magnetic multipole $\bOO{\text{mag}}{(l)}$ read,
\begin{equation} \label{Lomag}
\bs A_{\text{mag}}^{(l)}=-\dfrac{k^l}{4\,l!\,\varepsilon_0c} 
  \Bigl\{(l-1)\bs{u}\x\!\!\bigl[\bqq{(l)}{\text{mag}}\.\bs{h}_{\bs{u}}^{\,*}\bigr]
  +\bigl[\bs u\.\bs q_{\text{mag}}^{\,(l)}\bigr]\bs{e}_{\bs{u}}^{\,*}\Bigr\}.
\end{equation}
The first term in Eq.~(\ref{Jabapp}) thus becomes
\begin{equation}
\begin{array}{lll}
-c\doint_{S_\infty} \!\!\av{\bs J_{\text{mix}}}\td\sigma
&=-c\doint_{S_\infty}\bigl[\av{\bs L_{\text{mix}}}+\av{\bs S_{\text{mix}}}\bigr]\td\sigma
=\bs j_1+\bs j_2+\bs j_3
=\dsum_{l=1}^{\infty}\, \bs j_{\text{mix}}^{\,(l)}
\end{array}
\end{equation}
with
\begin{equation}
\begin{array}{lll}
\bs j^{\,(l)}_{\text{mix}}
&=\dfrac{k^{l-1}}{2\,l!}\,\,\text{Re}\doint_{4\pi}\!\!\td\Omega_u(-i)^{l+1}
\Bigl\{
\bs{e}_{\bs{u}}^{\,*}\x \bs q_{\text{elec}}^{\,(l)}+\dfrac1c\,\bs{h}_{\bs{u}}^{\,*}\x \bs q_{\text{mag}}^{\,(l)}
+(l-1)\bs u\x\Bigl[\dfrac1c\,
\bqq{(l)}{\text{mag}}\.\bs{h}_{\bs{u}}^{\,*}+
\bqq{(l)}{\text{elec}}\.\bs{e}_{\bs{u}}^{\,*}\Bigr]\Bigr\},
\end{array}
\end{equation}
which coincides with $\av{\bs T_{\text{mix}}^{(l)}}$ given in Eq. (\ref{Tmixint}), leading to the first equation of Eqs.~(29) in the main text,
\begin{eqnarray}
-c\doint_{S_\infty} \!\!\av{\bs J_{\text{mix}}}\td\sigma
=\av{\bs T_{\text{mix}}} .
\end{eqnarray}

The second term in Eq.~(\ref{Jabapp}), given by $(\bs j_4+\bs j_5)$ in Eq. (\ref{Jmixsca1}), involves only the scattered fields.
Based on Eqs.~(\ref{anbnapp}) and (\ref{alblapp}), 
the integral over solid angle can be performed to produce, after tedious algebra, the following relations
\begin{subequations}
\begin{eqnarray}
\doint_{4\pi}\!\!\td\Omega_u\,
\bs r\x\bigl[(\nabla\bs a_{\text{elec}}^{(l)})\.\bs a_{\text{elec}}^{(l')\,*}\bigr]
&=&\delta_{l,l'}\,\dfrac{k^{2l+4}}{4\pi\varepsilon_0^2}\,\dfrac{2^l(l^2+l-1)}{l\,(2l+1)!}\,
\bigl[\bOO{\text{elec}}{(l)}\tc{l-1}\bOO{\text{elec}}{(l)\,*}\bigr]\tc{2}\beps,\\[4mm]
\doint_{4\pi}\!\!\td\Omega_u\,
\bs r\x\bigl[(\nabla\bs a_{\text{mag}}^{(l)})\.\bs a_{\text{mag}}^{(l')\,*}\bigr]
&=&\delta_{l,l'}\,\dfrac{k^{2l+4}}{4\pi\varepsilon_0^2c^2}\,\dfrac{2^l(l^2+l-1)}{l\,(2l+1)!}\,
\bigl[\bOO{\text{mag}}{(l)}\tc{l-1}\bOO{\text{mag}}{(l)\,*}\bigr]\tc{2}\beps,
\end{eqnarray}
\begin{eqnarray}
\doint_{4\pi}\!\!\td\Omega_u\,
\bs r\x\bigl[(\nabla\bs a_{\text{elec}}^{(l)})\.\bs a_{\text{mag}}^{(l')\,*}\bigr]
&=&\dfrac{ik^{2l+3}}{4\pi\varepsilon_0^2c}\,\dfrac{2^{l}(l+1)}{(2l+1)!}\,
\bigl[\bOO{\text{elec}}{(l)}\tc{l-1}\bOO{\text{mag}}{(l-1)\,*}\bigr]\delta_{l-1,l'} \nonumber \\[4mm]  & & \mspace{15mu} {}
-\dfrac{ik^{2l'+3}}{4\pi\varepsilon_0^2c}\,\dfrac{2^{l'}(l'+1)}{(2l'+1)!}\,
\bigl[\bOO{\text{elec}}{(l)}\tc{l}\bOO{\text{mag}}{(l+1)\,*}\bigr]\delta_{l+1,l'} \\[4mm]
\doint_{4\pi}\!\!\td\Omega_u\,
\bs r\x\bigl[(\nabla\bs a_{\text{mag}}^{(l)})\.\bs a_{\text{elec}}^{(l')\,*}\bigr]
&=&
\dfrac{ik^{2l'+3}}{4\pi\varepsilon_0^2c}\,\dfrac{2^{l'}(l'+1)}{(2l'+1)!}\,
\bigl[\bOO{\text{mag}}{(l)}\tc{l}\bOO{\text{elec}}{(l+1)\,*}\bigr]\delta_{l+1,l'} \nonumber \\[4mm]  & & \mspace{15mu} {}
-\dfrac{ik^{2l+3}}{4\pi\varepsilon_0^2c}\,\dfrac{2^{l}(l+1)}{(2l+1)!}\,
\bigl[\bOO{\text{mag}}{(l)}\tc{l-1}\bOO{\text{elec}}{(l-1)\,*}\bigr]\delta_{l-1,l'}
\end{eqnarray}
\begin{eqnarray}
\doint_{4\pi}\!\!\td\Omega_u\,
\bigl[\bs a_{\text{elec}}^{(l)}\x\bs a_{\text{elec}}^{(l')\,*}\bigr]
&=&-\delta_{l,l'}\,\dfrac{k^{2l+4}}{4\pi\varepsilon_0^2}\,\dfrac{2^l}{l\,(2l+1)!}\,
\bigl[\bOO{\text{elec}}{(l)}\tc{l-1}\bOO{\text{elec}}{(l)\,*}\bigr]\tc{2}\beps\\[4mm]
\doint_{4\pi}\!\!\td\Omega_u\,
\bigl[\bs a_{\text{mag}}^{(l)}\x\bs a_{\text{mag}}^{(l')\,*}\bigr]
&=&-\delta_{l,l'}\,\dfrac{k^{2l+4}}{4\pi\varepsilon_0^2c^2}\,\dfrac{2^l}{l\,(2l+1)!}\,
\bigl[\bOO{\text{mag}}{(l)}\tc{l-1}\bOO{\text{mag}}{(l)\,*}\bigr]\tc{2}\beps
\end{eqnarray}
\begin{eqnarray}
\doint_{4\pi}\!\!\td\Omega_u\,
\bigl[\bs a_{\text{elec}}^{(l)}\x\bs a_{\text{mag}}^{(l')\,*}\bigr]
&=&\dfrac{ik^{2l+3}}{4\pi\varepsilon_0^2c}\,\dfrac{2^{l}(l+1)}{(2l+1)!}\,
\bigl[\bOO{\text{elec}}{(l)}\tc{l-1}\bOO{\text{mag}}{(l-1)\,*}\bigr]\delta_{l-1,l'} \nonumber \\[4mm]  & & \mspace{15mu} {}
-\dfrac{ik^{2l'+3}}{4\pi\varepsilon_0^2c}\,\dfrac{2^{l'}(l'+1)}{(2l'+1)!}\,
\bigl[\bOO{\text{elec}}{(l)}\tc{l}\bOO{\text{mag}}{(l+1)\,*}\bigr]\delta_{l+1,l'} \\[4mm]
\doint_{4\pi}\!\!\td\Omega_u\,
\bigl[\bs a_{\text{mag}}^{(l)}\x\bs a_{\text{elec}}^{(l')\,*}\bigr]
&=&
\dfrac{ik^{2l'+3}}{4\pi\varepsilon_0^2c}\,\dfrac{2^{l'}(l'+1)}{(2l'+1)!}\,
\bigl[\bOO{\text{mag}}{(l)}\tc{l}\bOO{\text{elec}}{(l+1)\,*}\bigr]\delta_{l+1,l'} \nonumber \\[4mm]  & & \mspace{15mu} {}
-\dfrac{ik^{2l+3}}{4\pi\varepsilon_0^2c}\,\dfrac{2^{l}(l+1)}{(2l+1)!}\,
\bigl[\bOO{\text{mag}}{(l)}\tc{l-1}\bOO{\text{elec}}{(l-1)\,*}\bigr]\delta_{l-1,l'}
\end{eqnarray}
\end{subequations}
Summing up, the second term in (\ref{Jabapp}) turns out to be
\begin{equation}
\begin{array}{lll}
-c\doint_{S_\infty} \!\!\av{\bs J_{\text{sca}}}\td\sigma
&=-c\doint_{S_\infty}\bigl[\av{\bs L_{\text{sca}}}+\av{\bs S_{\text{sca}}}\bigr]\td\sigma
=\bs j_4+\bs j_5=\dsum_{l=1}^{\infty}\, \bs j_{\text{sca}}^{\,(l)}
\end{array}
\end{equation}
with
\begin{equation}\label{jscal}
\av{\bs j_{\text{sca}}^{\,(l)}}
=-\dfrac{k^{2l+1}}{8\pi\varepsilon_0}\,\dfrac{2^l(l+1)}{(2l+1)!}\,\,\text{Im}
\bigl[\bOO{\text{elec}}{(l)}\tc{l-1}\bOO{\text{elec}}{(l)\,*}+\bOO{\text{mag}}{(l)}\tc{l-1}\bOO{\text{mag}}{(l)\,*}\bigr]\tc{2}\beps,
\end{equation}
which is identically $\av{\bs T_{\text{sca}}^{(l)}}$ in Eq. (\ref{Tscaapp}), so one eventually has
the second equation of Eq.~(29) in the main text,
\begin{equation}
-c\doint_{S_\infty} \!\!\av{\bs J_{\text{sca}}}\td\sigma=\av{\bs T_{\text{sca}}}.
\end{equation}

Finally we turn to the surface integral of the AM current density given by the right hand side of Eq. (3d) in the main text.
It is noted that the sum $\av{\bKKt}$ of the spin and orbital AM current densities,
$\av{\bKKs}$ and $\av{\bKKo}$,  does not give $\av{\bKK}=-\bs r\x\av{\bTT}$ that is usually termed
AM current density (AM flux tensor) \cite{zangwill,schwinger} and used for computing optical torque
based on the Maxwell stress tensor formalism \cite{jackson,zangwill,schwinger}, see, Eqs.~(\ref{tonrxT0}-\ref{tonrxT1}). That is
\begin{subequations}
\begin{equation}
\av{\bKKt}=\av{\bKKo}+\av{\bKKs}\ne\av{\bKK}=-\bs r\x\av{\bTT} \\
\end{equation}
but
\begin{equation}\label{tonKKt}
\oint_{S_\infty}\!\!\av{\bKK}\.\bs n\td\sigma
=\oint_{S_\infty}\!\!\bigl[\av{\bKKo}+\av{\bKKs}\bigr]\.\bs n\td\sigma
=\oint_{S_\infty}\!\!\av{\bKKt}\.\bs n\td\sigma,
\end{equation}
\end{subequations}
where $\av{\bTT}$ is the (symmetric) Maxwell stress tensor \cite{jackson,zangwill,schwinger},
the orbital AM current density $\av{\bKKo}$ and spin AM current density $\av{\bKKs}$ are given by
\begin{equation}
\begin{array}{ll}
\av{\bKKo}&=\dfrac{\varepsilon_0c^2}{4\omega}\text{Im}\,
\bigl[(\bs r\x\nabla\bs E)\x\bs B^*+
(\bs r\x\nabla\bs B^*)\x\bs E+\bs B^* \bs E +\bs E \bs B^*\bigr] \\[3mm]
\av{\bKKs}&=\dfrac{\varepsilon_0c^2}{2\omega}\text{Im}\,
\bigl[\bs B \bs E^* +\bs E^* \bs B-(\bs E^*\.\bs B)\bII\bigr]
\end{array}
\end{equation}
With the decomposition (\ref{ehtapp}), (\ref{ehincapp}), and (\ref{ehsapp}),
and taking advantage of Jones' lemma Eq. (\ref{jonesapp}), 
one has, after some algebra,
\begin{subequations}\label{j1j2j3j4j5a}
\begin{eqnarray}
\av{\bs T^{\,o}_{\text{mix}}}&=&-\doint_{S_\infty}\av{\bKKox}\.\bs n\td\sigma
=-c\doint_{S_\infty}\av{\bs L_{\text{mix}}}\td\sigma=\bs j_1+\bs j_2 ,\\ 
\av{\bs T^{\,s}_{\text{mix}}}&=&-\doint_{S_\infty}\av{\bKKsx}\.\bs n\td\sigma
=-c\doint_{S_\infty}\av{\bs S_{\text{mix}}}\td\sigma=\bs j_3, \\ 
\av{\bs T^{\,o}_{\text{sca}}}&=&-\doint_{S_\infty}\av{\bKKos}\.\bs n\td\sigma=-c\doint_{S_\infty}\av{\bs L_{\text{sca}}}\td\sigma=\bs j_4,\\
\av{\bs T^{\,s}_{\text{sca}}}&=&-\doint_{S_\infty}\av{\bKKss}\.\bs n\td\sigma=-c\doint_{S_\infty}\av{\bs S_{\text{sca}}}\td\sigma=\bs j_5,
\end{eqnarray}
\end{subequations}
which, together with
$$
\doint_{S_\infty}\bigl[\av{\bKKoi}+\av{\bKKsi}\bigr]\.\bs n\td\sigma=0 
$$
imply
\begin{equation}
\av{\bs T_{\text{mix}}}=\av{\bs T^{\,o}_{\text{mix}}}+\av{\bs T^{\,s}_{\text{mix}}} \quad \text{and} \quad
\av{\bs T_{\text{sca}}}=\av{\bs T^{\,o}_{\text{sca}}}+\av{\bs T^{\,s}_{\text{sca}}}
\end{equation}
and conclude the proof of Eq.~(\ref{tonKKt}) and thus Eq. (3d) in the main text.
In the derivation, use has been made of
$$
\oint_{S_\infty}\!\!\av{\bKKt}\.\bs n\td\sigma=\oint_{S}\av{\bKKt}\.\bs n\td\sigma,
$$
which follows the conservation of AM in free-space (see, also, discussion in Appendix \ref{appa}).

Equations~(\ref{j1j2j3j4j5a}) provide a physically transparent decomposition
for optical torque. The extinction torque due to orbital AM current, $\av{\bs T^{\,o}_{\text{mix}}}$,
the extinction torque coming from spin AM current, $\av{\bs T^{\,s}_{\text{mix}}}$,
the recoil torque stemming from
orbital AM current, $\av{\bs T^{\,o}_{\text{sca}}}$\revlina{, and}
the recoil torque arising from spin AM current, $\av{\bs T^{\,s}_{\text{sca}}}$,
are expressed, respectively, as follows
\begin{equation}
\begin{array}{ll}
\av{\bs T^{\,o}_{\text{mix}}}&=
\dsum_{l=1}^{\infty}\,
\bigl[\av{\bs T_{\text{mix}}^{(l)}}-\av{\bs T_{\text{mix}}^{\,s(l)}}
\bigr] \\[3mm]
\av{\bs T^{\,s}_{\text{mix}}}&=
\dsum_{l=1}^{\infty}\,
\av{\bs T_{\text{mix}}^{\,s(l)}} \\[5mm]
\av{\bs T^{\,o}_{\text{sca}}}
&=\dsum_{l=1}^{\infty}\,\Bigl[
\dfrac{(l^2+l-1)}{l(l+1)}\,\revlina{\av{\bs T_{\text{sca}}^{\,(l)}}}-\av{\bs j_{os}^{\,(l)}}\Bigr]
\\[5mm]
\av{\bs T^{\,s}_{\text{sca}}}
&=\dsum_{l=1}^{\infty}\,\Bigl[
\dfrac{1}{l(l+1)}\,\revlina{\av{\bs T_{\text{sca}}^{\,(l)}}}+\av{\bs j_{os}^{\,(l)}}\Bigr]
\end{array}
\end{equation}
where $\av{\bs T_{\text{mix}}^{(l)}}$ and
\revlina{$\av{\bs T_{\text{sca}}^{\,(l)}}$}
are given by (\ref{Tmixl}) and \revlina{(\ref{Tscal})}, 
 respectively, and
\begin{equation}
\begin{array}{ll}
\av{\bs T_{\text{mix}}^{\,s(l)}}=
 -\dfrac{1}{2\omega\,l!}\,\text{Im}\bigl[
 (\nabla^{(l)}\bs{E}_{\text{inc}}^*\,)\tc{l}\bOO{\text{mag}}{(l)}
 -(\nabla^{(l)}\bs{B}_{\text{inc}}^*\,)\tc{l}\bOO{\text{elec}}{(l)} \bigr] \\[5mm]
\av{\bs j_{os}^{\,(l)}}
=\dfrac{k^{2l}}{4\pi\varepsilon_0c}\,\dfrac{2^l(l+1)}{(2l+1)!}\,\,\text{Re}
\bigl[\bOO{\text{elec}}{(l)}\tc{l-1}\bOO{\text{mag}}{(l-1)\,*}-\bOO{\text{mag}}{(l)}\tc{l-1}\bOO{\text{elec}}{(l-1)\,*}\bigr],
\end{array}
\end{equation}
with the conventions $\bOO{\text{elec(mag)}}{\,(0)}=0$.
\revlina{
In the dipolar limit, similar decomposition was given in Refs. \cite{nieto2015a,nieto2015b,jiang2015}.}

\end{widetext}


\begin{thebibliography}{99}
\bibitem{antinieto} It is noted that Refs.\cite{nieto2015a,nieto2015b} presented
                   an expression for torque in the dipolar limit, which includes the recoil term.
                    However, it is inconsistent with this conclusion.
\end{thebibliography}
\end{document}